\documentclass[reprint,aps,graphicx,pre,showpacs]{revtex4-1}
\usepackage{amsmath,amssymb,amsfonts}     

\usepackage{graphicx,bm}
  \usepackage{paralist}
  \usepackage{epstopdf}
  \usepackage{graphics} 
 \usepackage[colorlinks=true]{hyperref}
 \hypersetup{urlcolor=blue, citecolor=red}
 \usepackage[latin1]{inputenc}
\usepackage[caption=false]{subfig}

\newcommand{\calU}{{\mathcal U}}

\newcommand{\calP}{{\mathcal P}}
\newcommand{\calQ}{{\mathcal Q}}

\newcommand{\calJ}{{\mathcal J}}

\newcommand{\R}{{\mathbb R}}

\renewcommand{\P}{\mathbb{P}}

\newcommand{\PP}{\widetilde{\mathcal P}}

\newcommand{\x}{\mathbf{x}}

\newcommand{\e}{{\mathrm e}}
\newcommand{\E}{{\mathbb E}}

\renewcommand{\P}{\mathbb P}
\newcommand{\p}{\widetilde{p}}

\newcommand{\q}{\widetilde{q}}
\newcommand{\J}{\widetilde{J}}

\newcommand{\U}{\widetilde{U}}

\begin{document}

 \title{Close encounters of the sticky kind: Brownian motion at absorbing boundaries}
 

\author{Paul C. Bressloff}
\address{Department of Mathematics, University of Utah 155 South 1400 East, Salt Lake City, UT 84112}

\keywords{Brownian motion, diffusion, absorption, Brownian local time}

\date{\today}

\begin{abstract} 
Encounter-based models of diffusion provide a probabilistic framework for analyzing the effects of a partially absorbing reactive surface, in which the probability of absorption depends upon the amount of surface-particle contact time. Prior to absorption, the surface is typically assumed to act as a totally reflecting boundary, which means that the contact time is determined by a Brownian functional known as the boundary local time. In this paper we develop a class of encounter-based models that deal with absorption at sticky boundaries. Sticky boundaries occur in a diverse range of applications, including cell biology, colloidal physics, finance, and human crowd dynamics. They also naturally arise in active matter, where confined active particles tend to spontaneously accumulate at boundaries even in the absence of any particle-particle interactions. We begin by constructing a one-dimensional encounter-based model of sticky Brownian motion (BM), which is based on the zero-range limit of non-sticky BM with a short-range attractive potential well near the origin. In this limit, the boundary-contact time is given by the amount of time (occupation time) that the particle spends at the origin. We calculate the joint probability density or propagator for the particle position and the occupation time, and then identify an absorption event as the first time that the occupation time crosses a randomly generated threshold. It follows that different models of absorption (Markovian and non-Markovian) correspond to different choices for the random threshold probability distribution. We illustrate the theory by considering diffusion in a finite interval with a partially absorbing sticky boundary at one end. We show how various quantities, such as the mean first passage time (MFPT) for single-particle absorption and the relaxation to steady-state at the multi-particle level, depend on moments of the random threshold distribution. Finally, we determine how sticky BM can be obtained by taking a particular diffusion limit of a sticky run-and-tumble particle (RTP). The latter is a canonical model for studying active matter in non-equilibrium physics and for modeling various processes in cell biology. The diffusion limit is well-defined provided that the velocity of the RTP is biased and spatially varying within a boundary layer around the origin. The heterogeneous component of the velocity plays an analogous role to the potential energy function of sticky BM. We thus establish encounter-based methods as a general framework for studying stochastic processes with absorption. \end{abstract}

\maketitle

\section{Introduction}

Sticky diffusion in a $d$-dimensional bounded domain is a stochastic process in which a diffusing particle can spend a finite amount of time ``stuck'' to the domain boundary. The sticking is usually assumed to be reversible, so an attached particle can re-enter the bulk domain. Whilst on the boundary, a particle may exhibit different dynamics from that observed in the interior (assuming $d>1$), that is, it may diffuse at a different rate. Sticky diffusion arises in a diverse range of applications. For example, in cell biology the exchange of molecules between the cytoplasm and cell membrane plays a major role in cell polarization and cell division \cite{Drake10,Jilkine11,Martin14,Halatek18}. A completely different application occurs in finance, where the evolution of interest rates exhibits sticky behavior near zero \cite{Longstaff92,Kabanov07}. Yet another example is the behavior of human crowds near walls \cite{Aurell20}. Sticky boundaries are also used to model interactions of colloidal particles in materials such as paint, toothpaste, and concrete \cite{Lu13}. Colloidal particles tend to interact attractively over length scales much smaller than their diameters, which have the range 100nm--10$\mu$m \cite{Manoharan15,Holmes17}. It is often observed that systems with short-range interactions are insensitive to the exact shape of the attractive potential well, depending mainly on the well depth and the well width. This has motivated a number of studies that treat the particles as diffusing hard spheres whose surfaces are sticky \cite{Baxter68,Fantoni06,Meng10,Holmes13}.

From a mathematical perspective, most work on sticky diffusion has been carried out within the probability community by considering sticky Brownian motion (BM) in one dimension (1D) \cite{Feller52,Ito63,Ito65,Karlin81,Ikeda89,Amir91,Engelbert14,Eberle19,Nguyen21}. Much of the focus of this work has been on determining properties of sticky BM within the context of weak and strong solutions of stochastic differential equations (SDEs), the theory of martingales, and alternative representations based on random walks in random environments. As far as we are aware, sticky diffusion processes have not received anything like the same level of attention within statistical physics. One possible reason for this is the singular nature of the probability measure in the sticky limit. Another obstacle has been the lack of methods for simulating sticky diffusion directly, which is related to the difficulty in ensuring that a particle hits the boundary exactly. Both of these issues have recently been addressed in Ref. \cite{Rabee20}. First, the authors show how sticky BM can arise as a limit of reflecting BM on the half-line with a strong short-range force at the origin. They establish this by performing an asymptotic expansion of the solution to the corresponding Fokker-Planck (FP) equation with respect to the width or inverse depth of the attractive potential. Second, they develop a numerical scheme based on discretizing space and constructing a Markov jump process whose generator locally approximates the generator of the sticky diffusion. (Note that the idea of modeling a sticky boundary in terms of a local potential well was previously proposed in Ref. \cite{Peters02}, but the connection to probabilistic notions of sticky BM were not explored. There have also been a number of complementary studies developing efficient numerical methods for implementing diffusion in the presence of partially reflecting (non-sticky) boundaries \cite{Andrews04,Erban07}.)

Another type of stochastic process where sticky boundary conditions arise is a velocity jump process. In particular, consider a particle moving on the half-line $(0,\infty)$ that switches between different constant velocity states according to a finite-state Markov chain. The particle hits the boundary at the origin whenever it is in a left-moving velocity state. A sticky boundary condition can be implemented by assuming that each time the particle hits the boundary, which is now a well-defined discrete event, it enters a bound state for an exponentially distributed random time interval before re-entering the interior of the domain. Velocity jump processes have a wide range of applications in cell biology \cite{Bressloff13}, including the run-and-tumble dynamics of bacteria \cite{Berg04,Angelani15,Angelani17}, the active motor-driven transport of vesicles within the interior of cells \cite{Newby10}, and the growth and shrinkage of polymer filaments during cell mitosis \cite{Dogterom93,Gopa11,Mulder12} and embryogenesis \cite{Bressloff19}. A sticky boundary could represent a bacterium temporarily attached to a container wall or a nucleation site for filament formation. Within the context of the non-equilibrium physics of active matter, a velocity jump process with a finite number of velocity states describes the motion of a so-called run-and-tumble particle (RTP) \cite{Volpe16}.
The implementation of a sticky boundary condition is much more straightforward for a velocity jump process compared to a diffusion process, due to the discrete nature of boundary collisions in the former case. 

A natural generalization of a sticky boundary is to include some form of surface absorption, so that the return from the boundary to the interior of the domain is not necessarily reversible. This could represent the removal of diffusing molecules in solution by an absorbent \cite{note1}. On the other hand, in the case of an RTP, absorption could represent the disappearance of a nucleation site for filament growth or the death of a bacterium attached to a wall. 
Recently, a general probabilistic framework for analyzing the effects of a partially absorbing reactive surface has been developed for diffusion processes in the case of non-sticky boundaries \cite{Grebenkov19,Grebenkov20,Grebenkov22,Bressloff22,Bressloff22a}. These so-called encounter-based models
take the probability of absorption to depend upon the amount of surface-particle contact time. Prior to absorption, the surface is assumed to act as a totally reflecting boundary, which means that the contact time is determined by a Brownian functional known as the boundary local time $\ell(t)$ \cite{Ito63,McKean75,Majumdar05}. (Although each boundary encounter takes place over an infinitely short time interval, in contrast to a sticky boundary, the particle returns to the boundary an infinite number of times in a finite interval before reentering the bulk domain, so that there is a measurable change in  the local time.) The local time also plays a key role in incorporating reflecting boundary conditions into stochastic differential equations (SDEs) for Brownian motion (BM). As we have shown elsewhere, it is also possible to develop an encounter-based model of a two-state RTP in the presence of a partially absorbing boundary, both in the sticky and non-sticky regimes \cite{Bressloff22rtp,Bressloff23}. Now the boundary contact time is the amount of time spent in the bound state at the origin. 

Encounter-based methods have two basic features that underly their utility and universality \cite{Grebenkov20,Bressloff22}: (i) The stochastic process of surface absorption is disentangled from the geometric structure of the bulk medium. This allows one to incorporate much more general absorption mechanisms than those with constant reactivity. (In the case of BM with constant surface reactivity, the corresponding FP equation involves a Robin or radiation boundary condition.). In particular, taking the probability of absorption to depend on the particle-surface contact time reflects the fact that a surface may need to be progressively activated by repeated encounters with a diffusing
particle, or an initially highly reactive surface may become less active due to multiple interactions with the particle, a process known as passivation \cite{Bartholomew01,Filoche08}. Alternatively, if absorption corresponds to the exit of a particle through a stochastically-gated ion channel or pore on the boundary of the domain, then it may have to return to the channel several times before it is open \cite{Bressloff15}. (ii)
A crucial step in the analysis of an encounter-based model is determining the joint probability density or generalized propagator for particle position and the particle-surface contact time. Once the propagator is known, absorption can be incorporated by identifying an absorption event as the first time that the particle-surface contact time crosses a randomly generated threshold. It follows that different models of absorption (Markovian and non-Markovian) correspond to different choices for the random threshold probability distribution. It turns out that the calculation of the propagator reduces to a two-step procedure. First, one solves a classical boundary value problem (BVP) for the probability density of particle position and a constant rate of absorption. Second, the absorption rate is reinterpreted as a Laplace variable $z$ conjugate to the surface-particle contact time, so that the propagator is obtained by evaluating the inverse Laplace transform of the classical solution with respect to $z$.

In this paper we develop an encounter-based model of sticky BM on the half-line and investigate the relationship between sticky BM and the diffusion limit of a run-and-tumble model. In both cases we proceed by generalizing the potential energy function method of Ref. \cite{Rabee20}. The structure of the paper is as follows. In Sect. II we describe two alternative formulations of sticky BM, a probabilistic version based on the slowing down of reflecting BM at the boundary \cite{Ito63} and the other based on the construction developed in Ref. \cite{Rabee20}. In the latter case, we include the details of the asymptotic analysis of the FP equation, since this will be extended in subsequent sections. In Sect. III, we present the encounter-based model of sticky BM with a partially absorbing boundary. First, we briefly summarize the corresponding model for non-sticky BM \cite{Grebenkov19,Grebenkov20,Bressloff22}, which involves the propagator $P(x,\ell,t)$ for the position $X(t)$ and boundary local time $\ell(t) $ at $x=0$. We then consider the propagator for reflecting BM in the presence of an attractive potential well that is localized to a small boundary layer $[0,\epsilon]$. Now the local time is replaced by an occupation time that specifies the amount of time the particle spends in the boundary layer up to time $t$. We then perform an asymptotic expansion of the solution to the propagator evolution equation along analogous lines to the FP equation, which allows us to derive a well-defined boundary condition for the propagator in the sticky limit $\epsilon \rightarrow 0$. The resulting occupation time $A_0(t)$ specifies the amount of time spent at the single point $x=0$, which by construction has non-zero probability or Lebesque measure. (One usually expects the probability at a single point to be zero.) It is related to the corresponding local time according to $A_0(t)=\nu \ell(t)/D$, where $\nu$ is a stickiness parameter \cite{Ito63} and $D$ is the diffusivity. Partial absorption is incorporated into the model in terms of a random occupation time threshold $\widehat{A}$ with distribution $\Psi(a)$, and the corresponding marginal density is $p^{\Psi}(x,t)=\int_0^{\infty} \Psi(a)P(x,a,t)da$. We also show how the encounter-based model of absorption at a non-sticky boundary is recovered in the limit $\nu \rightarrow 0$.

In Sects. IV and V we illustrate the theory by considering diffusion in the finite interval $[0,L]$ with a partially absorbing sticky boundary at $x=0$. First, we calculate the mean FPT (MFPT) for absorption at $x=0$ given a totally reflecting boundary at $x=L$. Taking an initial position $x_0\in (0,L)$, we show that the MFPT is $\tau(x_0)=[L^2-(L-x_0)^2]/2D + \E[a]L/\nu$, where $\E[a]$ is the mean occupation time (assuming that it is finite). Second, we establish the existence of a steady-state concentration for a population of independent Brownian particles with a constant influx at $x=L$, provided that $\E[a]<\infty$. We also show that the local rate of relaxation to the steady state depends on both the first and second moments $\E[a]$ and $\E[a^2]$.
Finally, in Sect. VI we explore the relationship between sticky BM and a sticky RTP on the half-line. We show that it is not possible to obtain sticky BM by taking the diffusion limit of the standard model of sticky run-and-tumble dynamics, where the particle enters and exits a bound state at $x=0$. However, the diffusion limit is well-defined if the velocity is biased and spatially varying within a boundary layer around the origin. The heterogeneous component of the velocity plays an analogous role to the potential energy function of sticky BM.  We thus establish encounter-based methods as a general framework for studying stochastic processes with absorption.

\section{Sticky Brownian motion in $\R^+$} 

The notion of sticky reflecting Brownian motion (BM) dates back to work by Feller in 1952, who considered various boundary conditions for the diffusion equation in $[0,\infty)$ that are consistent with stochastic processes behaving like standard Brownian motion in $(0,\infty)$ \cite{Feller52}. Ito and McKean \cite{Ito63} subsequently showed how to construct sample paths of sticky reflecting BM using a random time change that slowed down reflected paths so that the total time spent at the origin $x=0$ has positive Lebesgue measure. The latter is specified by the stickiness parameter. In this section we briefly describe the Ito and McKean construction and then consider the more recent formulation of sticky BM based on the inclusion of a strongly localized attractive potential close to the boundary at $x=0$ \cite{Rabee20}. 

\subsection{Slowed down reflecting BM}
Let $Y(t)$ be a reflecting BM, that is, $Y(t)=\sqrt{2D}|W(t)|$, where $W(t)$ is a standard Wiener process and $D$ is a constant diffusivity. That is,
\begin{equation}
\langle W(t)\rangle =0,\quad \langle W(t)W(t')\rangle =\min(t,t').
\end{equation}
Define the local time at the origin according to
\begin{equation}
\ell(t;Y)=\lim_{\epsilon \rightarrow 0} \frac{D}{\epsilon}\int_0^t {\mathbb I}_{[0,\epsilon]}(Y(\tau))d\tau,
\label{loc}
\end{equation}
where ${\mathbb I}_{\Sigma}(x)=1$ if $x\in \Sigma$ and is zero otherwise. It can be proven that $\ell(t)$ exists and is a positive, non-decreasing function of $t$. The corresponding SDE for $Y(t)$ is the so-called Skorokhod equation
\begin{equation}
dY(t)=\sqrt{2D}dW(t)+d\ell(t;Y).
\label{eqY}
\end{equation}
(Note that $\epsilon$, $Y(t)$ and $\ell(t)$ have units of length.) Formally speaking $d\ell(t;Y)=\delta(Y(t))$, which means that each time the particle hits the boundary it is given an impulsive kick back into the domain $x>0$.
Introduce the additive continuous increasing function
\begin{equation}
T(t)=t+\nu \ell(t;Y)/D,
\end{equation}
where $\nu$ is a fixed positive constant with units of length. Note that $dT/dt=1$ except at times where a trajectory contacts the origin. The reflected BM is slowed down at the origin by constructing the new stochastic process \cite{Ito63}
\begin{equation}
\widehat{Y}(t)=Y(T^{-1}(t)).
\end{equation}
The amount of time spent at the boundary $x=0$ can now be characterized in terms of the occupation time at the origin, as outlined in Ref. \cite{Karlin81}. The occupation time is defined as
\begin{equation}
\label{AY}
A(t;\widehat{Y}):=\int_0^t  {\mathbb I}_0(\widehat{Y}(s))ds=\int_0^t  {\mathbb I}_0(Y(T^{-1}(s))ds.
\end{equation}
Performing the change of variable for each sample path $s=T(\tau)$ gives
\begin{eqnarray}
&&A(t;\widehat{Y})=\int_0^{T^{-1}(t)}  {\mathbb I}_0({Y}(\tau))dT(\tau) \\
&=&\int_0^{T^{-1}(t)}  {\mathbb I}_0({Y}(\tau))d\tau+\frac{\nu}{D} \int_0^{T^{-1}(t)}  {\mathbb I}_0({Y}(\tau))d\ell(\tau;Y).\nonumber
\end{eqnarray}
The first integral on the right-hand side is simply the occupation time of reflected BM at a single point $x=0$ over the time interval $[0,T^{-1}(t)]$ and is identically zero. On the other hand, the second integral concentrates the integral to all points at which $ {\mathbb I}_0({Y}(\tau))=1$ so that $A(t)=\nu \ell(T^{-1}(t);Y)/D$. We thus obtain the result
\begin{equation}
\label{BCY}
A(t;\widehat{Y})=\frac{\nu}{D} \ell(t;\widehat{Y}).
\end{equation}
Clearly the amount of time the Brownian particle is stuck at the origin depends on the stickiness parameter $\nu$. Eqs. (\ref{eqY}) and (\ref{BCY}) correspond to the SDE for sticky reflecting BM $\widehat{Y}(t)$ \cite{note2}.

\subsection{Reflecting BM with a strong localized potential near the origin} 
\begin{figure}[t!]
\centering
\includegraphics[width=6cm]{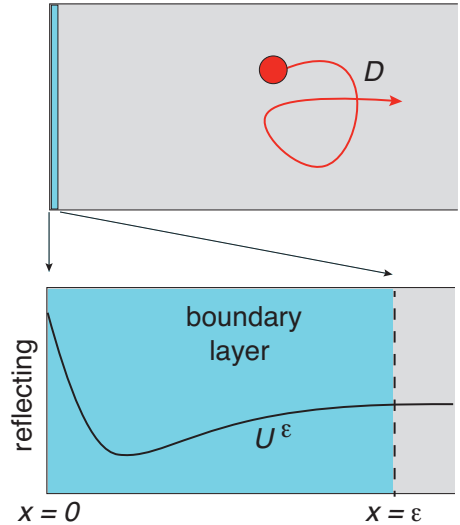}
\caption{Boundary layer construction near $x=0$ for sticky BM. Within the boundary layer $(0,\epsilon)$ there is a strongly attracting potential well, whereas the potential is approximately flat for $x\geq \epsilon$. }
\label{fig1}
\end{figure}

We now turn to a more recent formulation of sticky BM that is based on reflecting BM in $\R^+$ with a strongly localized attractive potential energy function in a neighborhood of the origin, as illustrated in Fig. \ref{fig1}. More specifically, let $U^{\epsilon}(x)\in C^2(\R^+)$ represent a family of potentials parameterized by $\epsilon$, $\epsilon >0$, with the following properties \cite{Rabee20}:
\begin{enumerate}

\item[(i)] $U^{\epsilon}(x)\approx 0$ for $x\geq \epsilon$. In particular, $U^{\epsilon}(x)$, $\partial_x U^{\epsilon(x)}$, $\partial_{xx}U^{\epsilon}(x) \leq O(\epsilon) $ for $x\geq \epsilon$. This assumption means that outside the boundary layer $(0,\epsilon)$, the force acting on the particle is negligible and it simply diffuses.

\item[(ii)] Within the boundary layer $(0,\epsilon)$ the function $U^{\epsilon}(x)$ takes the form of an attractive potential well that becomes deeper and narrower as $\epsilon \rightarrow 0$. The MFPT to escape from the boundary layer is approximately $\epsilon \e^{\Delta U}$ where $\Delta U$ is the height of the potential barrier.

\item[(iii)] There exists a parameter $\nu$ such that
\begin{equation}
\label{a3}
\lim_{\epsilon\rightarrow 0} \int_0^{\epsilon}\e^{-U^{\epsilon}(x)/k_BT}dx=\nu.
\end{equation}
This condition is the source of stickiness at the origin with $\nu$ ultimately being identified with the stickiness parameter.
Using steepest descents, it can be shown that
\begin{equation}
\sqrt{2\pi k_BT}\lim_{\epsilon\rightarrow 0}\frac{\e^{-U^{\epsilon}(x_0)/k_BT}}{\sqrt{\partial_{xx} U^{\epsilon}(x_0)}}=\nu,
\end{equation}
where $x_0$ is the location of the minimum of $U^{\epsilon}(x)$ in $(0,\epsilon)$. Assuming that $\partial_{xx}U^{\epsilon}(x_0)\approx U^{\epsilon}(x_0)/\epsilon^2$, it follows that $|U^{\epsilon}(x_0)|\sim |\log \epsilon|$ and the left-hand side is $O(1)$. We also see that the MFPT is proportional to $\nu$ so that increasing $\nu$ makes the boundary more sticky.
\end{enumerate}

\noindent An example of a family of potentials satisfying the above three properties is the Morse potential \cite{Rabee20}
\begin{equation}
U^{\epsilon}(x)=\overline{U}\left (1-\e^{-(x-x_0)/\xi}\right )^2-\overline{U},
\end{equation}
with $x_0= O(\epsilon^2)$, $\xi=O(\epsilon^2)$ and $\overline{U}$ defined implicitly as the solution to the equation   
\begin{equation}
\xi \e^{\overline{U}/k_BT}\sqrt{{\pi k_BT}/{\overline{U}}}= \nu.
\end{equation}
The latter is chosen so that Eq. (\ref{a3}) holds. 

 For any $\epsilon >0$, the position $X(t)$ of the particle evolves according to the SDE
\begin{equation}
\label{SDE}
 dX(t)=-\gamma^{-1}\partial_xU^{\epsilon}(X(t))dt+\sqrt{2D}dW(t),
\end{equation}
with a reflecting boundary at the origin. (More precisely, the stochastic variable should be written as $X_{\epsilon}(t)$ since the solution will depend on $\epsilon$. We drop the subscript for notational simplicity.)  The friction coefficient $\gamma$ satisfies the Einstein relation $\gamma D=k_BT$. Let $\rho(x,t)$ denote the probability density of the stochastic variable $X(t)$. The associated FP equation is given by
\begin{equation}
\label{FPBM}
\frac{\partial \rho}{\partial t}=D\frac{\partial^2 \rho }{\partial x^2}+\gamma^{-1}\frac{\partial }{\partial x}[\partial_xU^{\epsilon}(x)\rho],
\end{equation}
with $D\partial \rho(0,t)+\gamma^{-1}\partial_xU^{(\epsilon)}(0) \rho(0,t)=0$.

As shown in Ref. \cite{Rabee20}, one can use matched asymptotics to derive an effective FP equation in the limit $\epsilon \rightarrow 0$ that recovers the sticky boundary condition first introduced by Feller \cite{Feller52}. Since this analysis will be generalized to include the effects of absorption in Sect. III, we present the details here. Let $p(x,t)$ denote the leading order term in an asymptotic expansion of the outer solution in the region $x\geq \epsilon$, which evolves according to the standard diffusion equation
\begin{equation}
\label{pFP}
\frac{\partial p(x,t)}{\partial t}=D\frac{\partial^2p(x,t)}{\partial x^2},\quad x\geq \epsilon .
\end{equation}
Similarly let $q(x,t)$ denote the corresponding term of the inner solution with
\begin{equation}
\label{qFP}
\frac{\partial q(x,t)}{\partial t}=\gamma^{-1}\frac{\partial}{\partial x}\left [\partial_xU^{\epsilon}(x)q(x,t)\right ]+D\frac{\partial^2q(x,t)}{\partial x^2},
\end{equation}
for $x\in (0,\epsilon)$, together with the reflecting boundary condition
\begin{equation}
\label{qBC}
D \partial_x q(0,t)  +\gamma^{-1}\partial_x U^{\epsilon}(0) q(0,t)=0.
\end{equation}
Introducing the stretched coordinate $X=x/\epsilon$ and keeping only the leading order terms gives
\begin{equation}
0=\gamma^{-1}\frac{\partial}{\partial X}\left [\partial_XU^{\epsilon}(\epsilon X)q(\epsilon X,t)\right ]+D\frac{\partial^2q(\epsilon X,t)}{\partial X^2},
\end{equation}
for $X\in (0,1)$, with  $D\partial_Xq(0,t)+\gamma^{-1}\partial_XU^{\epsilon}(0)q(0,t)=0$. The solution for $q$ is
\begin{equation}
q(\epsilon X,t)={\mathcal N}(t)\e^{-U^{\epsilon}(\epsilon X)/k_BT},\quad X\leq 1,
\end{equation}
with the time-dependent amplitude ${\mathcal N}(t)$ determined by matching the inner solution $q(\epsilon X,t)$ with the outer solution $p(x,t)$ at $x=\epsilon$ and $X=1$.

Since the perturbation is not singular, matching can be performed at a single point, which is equivalent to imposing the continuity condition $p(\epsilon,t)=q(\epsilon,t)$. Using the fact that $\e^{-U^{\epsilon}(\epsilon)/k_BT} \sim 1$ and $p(\epsilon,t)\sim p(0,t)$, we obtain the leading order condition ${\mathcal N}(t)=p(0,t)$. A second condition on ${\mathcal N}(t)$ is obtained from the requirement that the total probability is conserved,
\begin{equation}
\frac{d}{\partial t}\left (\int_0^{\epsilon}q(x,t)dx+\int_{\epsilon}^{\infty} p(x,t)dx\right )=0.
\end{equation}
Moving the time derivatives inside the integrals and using the diffusion equation for $p$ yields
\begin{eqnarray}
0&=&\left (\int_0^{\epsilon}\frac{d{\mathcal N}(t)}{dt}\e^{-U^{\epsilon}(x)/k_BT}dx+D\int_{\epsilon}^{\infty} \frac{\partial^2 p(x,t)}{\partial x^2}dx\right )\nonumber \\
&=&\frac{d{\mathcal N}(t)}{dt}\int_0^{\epsilon} \e^{-U^{\epsilon}(x)/k_BT}dx -D\partial_xp(\epsilon,t).
\end{eqnarray}
Finally, taking the limit $\epsilon \rightarrow 0$ using Eq. (\ref{a3}) and setting ${\mathcal N}(t)=p(0,t)$ yields the boundary condition
\begin{equation}
\label{sBC1}
\nu\frac{\partial p(0,t)}{\partial t}=D\frac{\partial p(0,t)}{\partial x}.
\end{equation}
From the diffusion equation for $p$, we see that this is equivalent to the boundary condition first introduced by Feller \cite{Feller52}, namely,
\begin{equation}
\label{sBC2}
\nu\frac{\partial^2 p(0,t)}{\partial x^2}=\frac{\partial p(0,t)}{\partial x}.
\end{equation}
Formally speaking, the full probability density $\rho(x,t)$ can then be written in the form
\begin{equation}
\rho(x,t)= \lim_{\epsilon \rightarrow 0}p(x,t)\e^{-U^{\epsilon}(x)/k_BT} =p(x,t)(1+\nu \delta(x)).
\end{equation}
This follows from the assumed properties of $U^{\epsilon}(x)$. It will be convenient for our subsequent analysis to rewrite the full FP equation for sticky BM in the form
\begin{subequations}
\begin{eqnarray}
\label{FPea}
\frac{\partial p(x,t)}{\partial t}&=&D\frac{\partial^2p(x,t)}{\partial x^2},\quad x\in (0,\infty),\\
q(t)&=&\nu p(0,t),\quad \frac{dq(t)}{dt} =D\frac{\partial p(0,t)}{\partial x}.
\label{FPeb}
\end{eqnarray}
\end{subequations}
This makes explicit the notion that the probability of being at the origin has finite Lebesgue measure and consequently a nonzero occupation time. Note that if $\nu=0$ then $q(t)=0$ and we recover reflecting BM.

\begin{figure*}[t!]
\centering
\includegraphics[width=16cm]{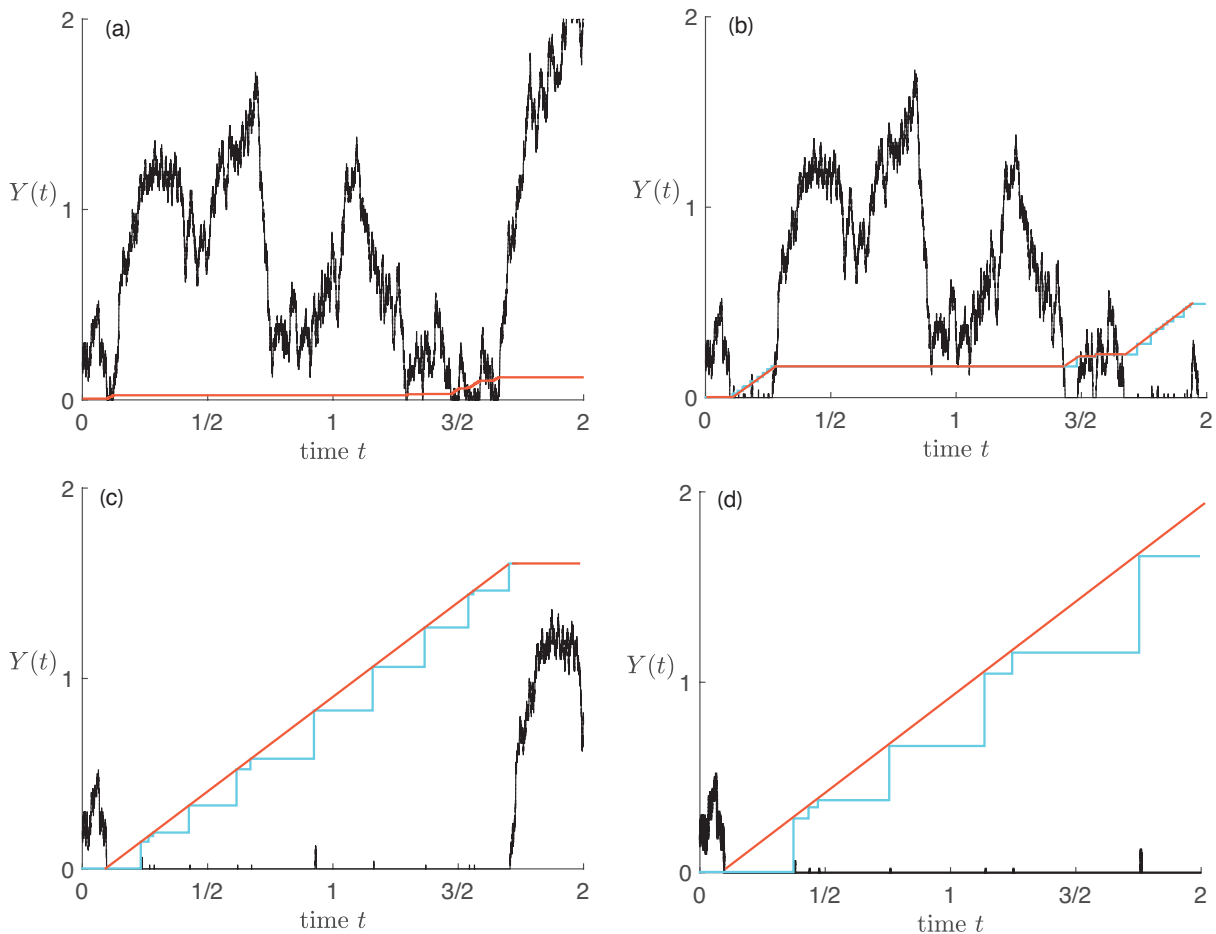}
\caption{Sample paths $Y(t)$ of the sticky random walk algorithm presented in Ref. \cite{Rabee20} for different values of the stickiness parameter: (a) $\nu=0.1$; (b) $\nu=1$; (c) $\nu=10$; (d) $ \nu=20$. Other parameter values are $x_0=0.2$ and $h=0.01$. We also plot the accumulation time $A_0(t)$ at the origin, which is shown as a piecewise smooth (red) curve. Short-lived excursions away from the origin are smoothed out. However, the time intervals between such excursions are indicated by the (blue) staircase). }
\label{fig2}
\end{figure*}

\subsection{Numerical method for simulating sticky BM} The derivation of the FP equation for sticky BM, see Eqs. (\ref{FPea}) and (\ref{FPeb}), was based on taking the limit $\epsilon \rightarrow 0$ of reflecting BM in the presence of a local attractive potential well of width $\epsilon$. This will be the foundation of our encounter-based model of sticky BM developed analytically in subsequent sections. However, as highlighted in Ref. \cite{Rabee20}, simulating the corresponding SDE (\ref{SDE}) for small $\epsilon$ in order to generate sample paths of sticky BM is not particularly efficient, since a very small step size is needed in order to achieve reasonable accuracy. Here we briefly describe a more effective numerical scheme for simulating sticky BM that was also introduced in Ref. \cite{Rabee20}. This  is based on constructing a continuous time Markov chain that represents a sticky random walk. We use this algorithm to illustrate the fact that the occupation time at the origin, see Eq. (\ref{BCY}), has a non-zero Lebesgue measure. 

Suppose that $\R^+$ is discretized by setting $x=nh$ for positive integers $n$ with $h$ the lattice spacing. The sticky random walk on the lattice can be simulated exactly using a simple Monte Carlo algorithm. Let $N(t)$ denote the lattice site occupied by the random walker at time $t$. The position is updated as follows \cite{Rabee20}:
\medskip

\noindent (I) If $N(t)\geq 1$ then the particle jumps to one of the neighboring lattice sites $N(t)\pm 1$ with equal probability. The waiting time for the jump is exponentially distributed with mean waiting time $h^2/2$.
\medskip

\noindent (II) If $N(t)=0$ then the particle jumps to the right with unit probability. The waiting time for the jump is again exponentially distributed but the mean waiting time is now $h^2/2+\nu h$, where $\nu$ is the stickiness parameter.

\medskip

\noindent Example realizations of the sticky random walk are shown in Fig. \ref{fig2}. The plots are generated using the MatLab code presented in appendix A of Ref. \cite{Rabee20}, which has been slightly modified in order to display the occupation time $A_0(t)$ at the origin. As expected the occupation time tends to increase with $\nu$.

\setcounter{equation}{0}

\section{Encounter-based model of absorption} 

In this section we develop an encounter-based model of partial absorption at a sticky boundary ($\nu >0$). We begin by briefly reviewing the case of a non-sticky boundary ($\nu=0$), which has been analyzed elsewhere \cite{Grebenkov20,Grebenkov22,Bressloff22,Bressloff22a}. 

\subsection{Partial absorption at a non-sticky boundary and the local time propagator}
The classical Robin boundary condition for partially reflecting BM in $\R^+$ takes the form
\begin{equation}
\label{Robin}
D\partial_xp(0,t)=\overline{\kappa}_0 p(0,t),
\end{equation}
where $\overline{\kappa}_0$ is a positive constant absorption rate or reactivity parameter. Note that $\overline{\kappa}_0$ has units of speed. Recently, it has been shown how to reformulate the Robin boundary condition for a diffusing particle using a probabilistic interpretation based on the boundary local time (\ref{loc}) \cite{Grebenkov20,Grebenkov22,Bressloff22,Bressloff22a}. Let $P(x,\ell,t)$ denote the joint probability density or local time propagator for the pair $(X(t),\ell(t))$.
It can be proven that the propagator $P(x,\ell,t)$ evolves according to the equations
\cite{Grebenkov20,Bressloff22}
\begin{subequations}
\begin{eqnarray}
\label{Ploc1}
  \frac{\partial P(x,\ell,t)}{\partial t}&=&D\frac{\partial^2 P(x,\ell,t)}{\partial x^2},\ x\in (0,\infty),\\
\label{Ploc2}
  \partial_xP(0,\ell,t)&=& \delta(\ell)  P(x,0,t)  +D\frac{\partial}{\partial \ell} P_0(0,\ell,t).
\end{eqnarray}
\end{subequations}
Since the local time only changes at the boundary, the propagator satisfies the standard diffusion equation away from the boundary. The boundary condition at $x=0$ can be interpreted as a form of probability conservation  whereby a net flux into the boundary leads to a shift in the joint density to larger values of $\ell$. The Dirac delta function ensures that if $\ell(t)=0$ then $P(x,\ell,t)=0$ for all $\ell>0$.

 In order to incorporate partial absorption at $x=0$, we introduce the stopping time 
\begin{equation}
\label{Tell0}
{\mathcal T}=\inf\{t>0:\ \ell(t) >\widehat{\ell}\},
\end{equation}
 with $\widehat{\ell}$ a random variable that represents a local time threshold with $\P[\widehat{\ell}>\ell]=\Psi(\ell)$. The stopping time ${\mathcal T}$ is the FPT for the event that $\ell(t)$ crosses the random threshold $\widehat{\ell}$, which we identify with the time at which absorption occurs.  The relationship between the marginal probability density $p^{\Psi}(\x,t)$ for a given distribution $\Psi$ and $P(\x,\ell,t)$ can then be established by noting that
\[p(x,t )dx=\P[X(t) \in (x,x+dx), \ t < {\mathcal T}].\]
Since $\ell(t)$ is a nondecreasing process, the condition $t < {\mathcal T}$ is equivalent to the condition $\ell(t) <\widehat{\ell}$. 
This implies 
\[p^{\Psi}(x,t)dx=\P[X(t) \in (x,x+dx),\  \ell(t) < \widehat{\ell}], \]
and hence
\begin{eqnarray}
 p^{\Psi}(x,t) 
&=&\int_0^{\infty} d\ell \ \psi(\ell)\int_0^{\ell} d\ell' P(x,\ell',t)\nonumber \\
&=&\int_0^{\infty}\Psi(\ell) P (x,\ell,t)d\ell , 
\label{peep}
\end{eqnarray}
after reversing the order of integration. Note that $\psi(\ell)=-\Psi'(\ell)$.

The crucial step in the above encounter-based model of absorption is the observation that if $\Psi(\ell)=\e^{-\gamma \ell}$ with $\gamma=\overline{\kappa}_0/D$, then the marginal density $p^{\Psi}(x,t)$ evolves according to the diffusion equation with the Robin boundary condition (\ref{Robin}) at $x=0$. Since $\Psi(\ell)=\e^{-\gamma \ell}$, we see that the solution to the Robin BVP is the Laplace transform of the local time propagators $P(x,\ell,t)$ with respect to the local time $\ell$, and $\gamma$ can be identified as the conjugate Laplace variable $z$. That is, (on dropping the superscript $\Psi$)
\begin{eqnarray}
p(x,t)|_{\kappa_0=zD}=\widetilde{P}(x,z,t):=\int_0^{\infty}\e^{-z\ell}P(x,\ell,t)d\ell.\nonumber \\
\end{eqnarray}
Assuming that we can invert the Laplace transform, the marginal density for a general probability distribution $\Psi(\ell) $ can be found by setting
     \begin{eqnarray}
     \label{int}
  p^{\Psi} (x,t)&=\int_0^{\infty} \Psi(\ell){\mathcal L}_{\ell}^{-1}[\widetilde{P}(x,z,t)]d\ell .
  \end{eqnarray}
  where ${\mathcal L}^{-1}$ denotes the inverse Laplace transform.
  In conclusion, a general model of partial absorption for BM with a partially reflecting boundary involves solving the Robin BVP, reinterpreting the constant absorption rate $\overline{\kappa}_0$ in terms of the Laplace variable $z=\overline{\kappa}_0/D$ that is conjugate to the local time, inverting the Laplace transform with respect to $z$, and then evaluating the integral (\ref{int}).

\subsection{Partial absorption at a sticky boundary and the occupation time propagator} 

Now suppose that $\nu >0$. The asymptotic construction of Ref. \cite{Rabee20}, which was summarized in Sect. IIB, can be adapted to incorporate partial absorption when the boundary at $x=0$ is sticky. For any $\epsilon >0$, we define the occupation time within the boundary layer to be the Brownian functional
\begin{equation}
A_{\epsilon}(t)=\int_0^t{\mathbb I}_{[0,\epsilon]}(X(\tau))d\tau,
\end{equation}
where $X(t)$ evolves according to Eq. (\ref{SDE}). As we have previously shown \cite{Bressloff22,Bressloff22a}, one can formulate an encounter-based model of absorption within the spatially extended subdomain $[0,\epsilon]$ by considering the corresponding occupation time propagator 
\begin{eqnarray}
&& \calP(x,a,t)dxda\nonumber \\
 && =\P[x<X(t)<x+dx,a<A_{\epsilon}(t)<a+da],
\end{eqnarray}
whose evolution equation is
\begin{subequations}
\begin{eqnarray}
 &&\frac{\partial \calP(x,a,t)}{\partial t}=\gamma^{-1}\frac{\partial}{\partial x}\left [\partial_xU^{\epsilon}(x)\calP(x,a,t)\right ]+D\frac{\partial^2\calP(x,a,t)}{\partial x^2}\nonumber \\
 \label{BMsticka}
 \end{eqnarray}
 for $ x\geq \epsilon$ and
 \begin{eqnarray}
 &&\frac{\partial \calP(x,a,t)}{\partial a}+\frac{\partial \calP(x,a,t)}{\partial t}+\delta(a)\calP(x,0,t)\\
 &&\quad =\gamma^{-1}\frac{\partial}{\partial x}\left [\partial_xU^{\epsilon}(x)\calP(x,a,t)\right ]+D\frac{\partial^2\calP(x,a,t)}{\partial x^2}\nonumber 
\label{BMstickb}
\end{eqnarray}
for $ x\in (0,\epsilon)$, together with a reflecting boundary condition at $x=0$,
\begin{equation}
D\partial_x\calP(0,a,t) +\gamma^{-1} \partial_x U^{\epsilon}(0) \calP(0,a,t),
\end{equation}
and the continuity conditions
\begin{equation}
\label{cont0}
P(\epsilon^-,a,t)=\calP(\epsilon^+,a,t),\quad \partial_xP(\epsilon^-,a,t)=\partial_xP(\epsilon^+,a,t).
\end{equation}
\end{subequations}
Eq. (\ref{BMstickb}) takes the form of an age-structured model, reflecting the fact that whenever the particle is within the boundary layer, the occupation time $A_{\epsilon}(t)$ increases at the same rate as the absolute time $t$. (Age-structured models are typically found within the context of birth-death processes in ecology and cell biology, where the birth and death rates of individual organisms and cells depend on their age
\cite{McKendrick25,Foerster59,Iannelli17}.) The term involving the Dirac delta function $\delta(a)$ ensures that the probability of being in the boundary layer is zero if the occupation time is zero.

Let $P(x,a,t)$ be the leading order term in an asymptotic expansion of the outer solution for $x\geq \epsilon$, which evolves according to the standard diffusion equation  
\begin{subequations}
\begin{eqnarray}
\frac{\partial  P(x,a,t)}{\partial t}=D\frac{\partial^2 P(x,a,t)}{\partial x^2},\quad x \geq \epsilon .\end{eqnarray}
Similarly, let $Q(x,a,t)$ to be the corresponding leading order term in the asymptotic expansion of the inner solution:
\begin{eqnarray}
 &&\frac{\partial Q(x,a,t)}{\partial a}+\frac{\partial Q(x,a,t)}{\partial t}+\delta(a)Q(x,0,t)\\
 && \quad =\gamma^{-1}\frac{\partial}{\partial x}\left [\partial_xU^{\epsilon}(x)Q(x,a,t)\right ] 
  +D\frac{\partial^2Q(x,a,t)}{\partial x^2}\nonumber 
  \end{eqnarray}
for $ x\in (0,\epsilon)
$, and the reflecting boundary condition
\begin{equation}
D\partial_xQ(0,a,t) +\gamma^{-1} \partial_x U^{\epsilon}(0) Q(0,a,t)=0.
\end{equation}
The continuity conditions at $x=\epsilon$ become
\begin{equation}
\label{cont}
P(\epsilon,a,t)=Q(\epsilon,a,t),\quad \partial_xQ(\epsilon,a,t)=\partial_xP(\epsilon,a,t).
\end{equation}
\end{subequations}

\begin{figure*}[t!]
\centering
\includegraphics[width=16cm]{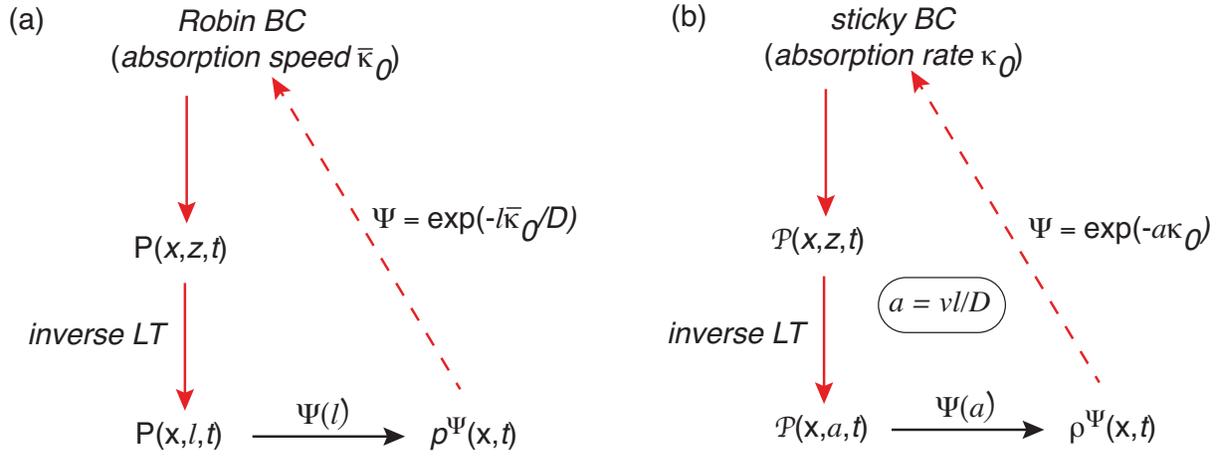}
\caption{Diagrams illustrating the encounter-based framework for BM motion on the half-line with a partially absorbing boundary at the origin. (a) Non-sticky boundary. The solution of the diffusion equation for the Robin boundary condition (BC) with constant reactivity $\overline{\kappa}_0$ generates the Laplace transform (LT) $\widetilde{P}(x,z,t)$ of the local time propagator $P(x,\ell,t)$ with $z=\overline{\kappa}_0/D$. The density $p^{\Psi}(\x,t)$ is given by Eq. (\ref{int}). (b) Sticky boundary. The solution of the diffusion equation for the sticky BC (\ref{pnub}) and (\ref{pnuc}) in the case of a constant reactivity $\kappa_0$ generates the Laplace transform $\widetilde{\calP}(x,z,t)$ of the occupation time propagator $\calP(x,a,t)$ with $z=\kappa_0$. The singular density is $\rho^{\Psi}(x,t)=(1+\nu\delta(x))p^{\Psi}(x,t)$ with $p^{\Psi}$ given by Eq. (\ref{ppp}) and $\nu$ the stickiness parameter.}
\label{fig3}
\end{figure*}

The matched asymptotics proceeds along analogous lines to Sect. IIB. Introducing the stretched coordinate $X=x/\epsilon$ and keeping only the leading order terms we find that
\begin{equation}
\label{Qeps}
Q(\epsilon X,a,t)={\mathcal N}(a,t)\e^{-U^{\epsilon}(\epsilon X)/k_BT},\quad X\leq 1,\ a>0.
\end{equation}
The first continuity condition in (\ref{cont}) implies that to leading order ${\mathcal N}(a,t)=P(0,a,t)$. Next, substituting the solution (\ref{Qeps}) into Eq. (\ref{BMstickb}) and integrating with respect to $x\in (0,\epsilon)$ gives
\begin{eqnarray}
 &&\left [\frac{\partial {\mathcal N}(a,t))}{\partial a}+\frac{\partial {\mathcal N}(a,t)}{\partial t}+\delta(a){\mathcal N}(0,t)\right ]\int_0^{\epsilon}\e^{-U^{\epsilon}(x)/k_BT}dx \nonumber\\
 && =\gamma^{-1} \left [\partial_xU^{\epsilon}(\epsilon)Q(\epsilon,a,t)-\partial_xU^{0}(\epsilon)Q(0,a,t)\right ]\nonumber \\
 &&\quad +D\partial_xQ(\epsilon,a,t)-D\partial_xQ(0,a,t).
\end{eqnarray}
Imposing the reflecting boundary condition at $x=0$ and taking the limit $\epsilon \rightarrow 0$ using Eq. (\ref{a3}) finally yields the propagator boundary condition
\begin{eqnarray}
 &&\nu\left [\frac{\partial P(0,a,t))}{\partial a}+\frac{\partial P(0,a,t)}{\partial t}+\delta(a)P(0,0,t)\right ]\nonumber \\
 &&\qquad =D\partial_xP(0,a,t).
\end{eqnarray}
In summary, the full propagator has the formal solution
\begin{equation}
\calP(x,a,t)=P(x,a,t)(1+\nu \delta(x)),
\end{equation}
with
\begin{subequations}
\begin{eqnarray}
\label{BVPa}
&&\frac{\partial  P(x,a,t)}{\partial t}=D\frac{\partial^2 P(x,a,t)}{\partial x^2},\quad x >0,\\
&&\label{BVPb} Q(a,t)=\nu P(0,a,t),\\
&& \left [\frac{\partial Q(a,t))}{\partial a}+\frac{\partial Q(a,t)}{\partial t}+\delta(a)Q(0,t)\right ]=D\partial_xP(0,a,t).\nonumber \\
\label{BVPc}
 \end{eqnarray}
\end{subequations}

We can interpret $P(x,a,t)$ as the propagator for the occupation time $A_0(t)$ at the single point $x=0$:
\begin{equation}
A_{0}(t)=\lim_{\epsilon \rightarrow 0}\int_0^t{\mathbb I}_{[0,\epsilon]}(X(\tau))d\tau.
\end{equation}
Partial absorption at the sticky end $x=0$ can now be introduced along analogous lines to a non-sticky boundary by imposing the stopping condition 
\begin{equation}
\label{TA}
{\mathcal T}=\inf\{t>0:\ A_0(t) >\widehat{A}\},
\end{equation}
with $\P[\widehat{A}>a]=\Psi(a)$, and setting
\begin{eqnarray}
 p^{\Psi}(x,t) &=&\int_0^{\infty}\Psi(a) P(x,a,t)da,\\ q^{\Psi}(t)&=&\nu \int_0^{\infty}\Psi(a) P(0,a,t)da . 
\label{peerho}
\end{eqnarray}
In the case of the exponential distribution $\Psi(a)=\e^{-\kappa_0 a}$, we obtain the diffusion equation
for a sticky partially absorbing boundary (after dropping the superscript $\Psi$)
\begin{subequations}
\begin{eqnarray}
\label{pnua}
\frac{\partial p(x,t)}{\partial t}&=&D \frac{\partial^2 p(x,t)}{\partial x^2},\ x >0,\\
\label{pnub}
 q(t)&=&\nu p(0,t), \\
\frac{d q(t)}{d t}&=&-\kappa_0 q(t)+D\frac{\partial p(0,t)}{\partial x} .
\label{pnuc}
\end{eqnarray}
\end{subequations}
(Note that $\kappa_0$ has units of inverse time.). Given the solution to Eqs. (\ref{pnua})--(\ref{pnub}), we reinterpret $\kappa_0$ as the Laplace variable conjugate to the occupation time by setting $\widetilde{P}(x,z,t)=p(x,t)|_{\kappa_0=z}$ and $\widetilde{Q}(z,t)=q(t)|_{\kappa_0=z}$ so that
\begin{subequations}
\begin{eqnarray}
\label{ppp}
  p^{\Psi}(x,t) &=&\int_0^{\infty}\Psi(a) {\mathcal L}^{-1}_a [P(x,z,t)]da ,\\  q^{\Psi}(t)
 &=&\int_0^{\infty}\Psi(a) {\mathcal L}^{-1}_a [Q(z,t)]da.
\label{pho2}
\end{eqnarray}
\end{subequations}

Comparing the boundary conditions (\ref{pnub}) and (\ref{pnuc}) with the Robin boundary condition (\ref{Robin}) implies that in order to recover the latter in the limit $\nu \rightarrow 0$, it is necessary to simultaneously take the limit $\kappa_0\rightarrow \infty$ such that $\overline{\kappa}_0=\nu\kappa_0$ is fixed. (If we only took the limit $\nu \rightarrow 0$ then the boundary would become totally reflecting.) Now recall the relationship (\ref{BCY}) between the occupation time and local time for sticky BM. Setting $a=\nu \ell/D$ in Eqs. (\ref{peerho}) implies that
\begin{subequations}
\begin{eqnarray}
p^{\Psi}(x,t) &=&\frac{\nu}{D}\int_0^{\infty}\Psi(\nu \ell/D) P(x,\nu \ell/D,t)d\ell,\nonumber \\\\ q^{\Psi}(t)&=&\frac{\nu^2}{D} \int_0^{\infty}\Psi(\nu \ell/D) P(0,\nu \ell/D,t)d\ell . \nonumber \\
\end{eqnarray}
\end{subequations}
Hence, define the local time propagator according to
\begin{equation}
\P_{\rm loc}(x,\ell,t)=\frac{\nu}{D}P(x,\nu \ell/D,t),
\end{equation}
and set $\Psi_{\rm loc}(\ell)=\Psi(\nu \ell/D)$. Assuming that these are both well-defined in the limit $\nu \rightarrow 0$ (by an appropriate scaling of parameters consistent with fixed $\overline{\kappa}_0=\nu \kappa_0$), we find that
\begin{equation}
\lim_{\nu\rightarrow 0}p^{\Psi}(x,t)=\int_0^{\infty}\Psi_{\rm loc}(  \ell) P_{\rm loc}(x,  \ell,t)d\ell,\ \lim_{\nu\rightarrow 0}q^{\Psi}(t)=0.
\end{equation}
That is, we recover the encounter-based model of partially reflecting BM considered in Sect. IIIA The encounter-based schemes for sticky and non-sticky BM are summarized in Fig. \ref{fig3}

\setcounter{equation}{0}
\section{First passage time problem}

In order to illustrate the encounter-based formulation of sticky BM, consider a particle diffusing in a finite interval $[0,L]$ with a partially absorbing, sticky boundary at $x=0$ and a reflecting boundary at $x=L$. We are interested in calculating the MFPT for absorption. For a constant absorption rate $\kappa_0$, the probability density evolves according to the initial BVP given by Eqs. (\ref{pnua}) and (\ref{pnub}) together with the reflecting boundary condition at $x=L$, $\partial_xp(L,t)=0$. We will solve these equations by Laplace transforming with respect to time $t$ under the initial condition $p(x,0)=\delta(x-x_0)$ with $0<x_0<L$.  That is, making the initial position explicit and setting $\p(x,s|x_0)=\int_0^{\infty}\e^{-st}p(x,t|x_0)dt$ and $\q(s|x_0)=\int_0^{\infty}\e^{-st}q(t|x_0)dt$ with $\q(s|x_0)=\nu \p(0,s|x_0)$, we have
\begin{subequations}
\begin{eqnarray}
\label{pnuLTa}
D \frac{\partial^2 \p(x,s|x_0)}{\partial x^2}-s\p(x,s|x_0)&=&-\delta(x-x_0),\\
\label{pnuLTc}
(s+\kappa_0) \p(0,s|x_0)&=&\frac{D}{\nu }\frac{\partial \p(0,s|x_0)}{\partial x} ,\\ 
\frac{\partial \p(L,s|x_0)}{\partial x}&=&0.
\label{pnuLTc}
\end{eqnarray}
\end{subequations}
A solution of Eq. (\ref{pnuLTa}) for $0\leq x<x_0$ that satisfies the boundary condition at $x=0$ takes the form
\begin{subequations}
\begin{eqnarray}
\label{pl}
\p_<(x,s)
 &=&\frac{\sqrt{sD}\cosh(\sqrt{s/D}x)+\nu(s+\kappa_0)\sinh(\sqrt{s/D}x)}{\sqrt{sD}+\nu(s+\kappa_0)}.\nonumber\\
\end{eqnarray}
Similarly, a solution of Eq. (\ref{pnuLTa}) for $L>x>x_0$ that satisfies the boundary condition at $x=L$ is
\begin{equation}
\label{pg}
\p_>(x,s)= \cosh(\sqrt{s/D }(L-x)).\end{equation}
\end{subequations}
Imposing continuity of the solution across $x=x_0$ and the flux discontinuity condition $ \partial_x\p(x_0^+,s)- \partial_x\p(x_0^-,s)=-1/D$ implies that
\begin{equation}
\p(x,s|x_0)=\left \{ \begin{array}{cc} A(s)\p_<(x,s)\p_>(x_0,s),& 0\leq x < x_0 \\
A(s)\p_<(x_0,s)\p_>(x,s),& x< x_0 \leq L, \end{array} \right .
\label{psticky}
\end{equation}
with
\begin{eqnarray}
A(s)&=&\frac{\sqrt{sD}+\nu(s+\kappa_0)}{\sqrt{sD }} \Lambda(s),\\
\Lambda(s)&=&\frac{1}{\sqrt{sD}\sinh(\sqrt{s/D}L)+\nu(s+\kappa_0)\cosh(\sqrt{s/D}L)}.\nonumber
\end{eqnarray}

The survival probability at time $t$ is
\begin{equation}
\label{SP}
S(x_0,t)=\int_0^{L}\rho(x,t|x_0)dx=\int_0^{L}p(x,t|x_0)dx+q(t|x_0).
\end{equation}
Note that there is a non-zero contribution from the probability that the particle is at the origin, reflecting the stickiness of the boundary.
Differentiating both sides of Eq. (\ref{SP}) with respect to $t$ using Eqs. (\ref{pnuLTa}) and the boundary conditions, we have
\begin{eqnarray}
 \frac{\partial S(x_0,t)}{\partial t}&=&D\int_0^L\frac{\partial^2 \rho(x,t|x_0)}{\partial x^2}dx+\frac{dq(t|x_0)}{dt}\nonumber \\
 &=&-D\frac{\partial \rho(0,t|x_0)}{\partial x}+\frac{dq(t|x_0)}{dt}\nonumber \\
 &=&- \kappa_0 q(t|x_0).
\end{eqnarray}
Let $T(x_0)$ denote the FPT for absorption at $x=0$. The FPT density is
\begin{equation}
f(x_0,t):=-\frac{\partial S(x_0,t)}{\partial t}=\kappa_0 q(t|x_0),
\end{equation}
and the corresponding MFPT is
\begin{eqnarray}
\tau(x_0):= \E[ T(x_0)] &=& \kappa_0\int_0^{\infty}t q(t|x_0)dt\nonumber \\
&=&-\nu  \kappa_0 \lim_{s\rightarrow 0} \frac{\partial}{\partial s}\p(0,s|x_0).
  \label{MFPTa}
\end{eqnarray}
Setting $x=0$ in the solution (\ref{psticky}) gives
\begin{eqnarray}
\label{px0}
&&\p(0,s|x_0)=\Lambda(s)\ \cosh(\sqrt{s/D}(L-x_0)). 
\end{eqnarray}
We thus find that
\begin{equation}
\tau(x_0)=\frac{L^2-(L-x_0)^2}{2D}+\frac{L+\nu}{\nu \kappa_0}.
\end{equation}
The first term on the right-hand side is the MFPT for diffusion with a totally absorbing boundary at $x=0$, whereas the second term is the contribution from the joint effects of stickiness and partial absorption. 
    
Having determined $\p(x,s|x_0)$ and $\tau(x_0)$ for a constant rate of absorption, we can now consider a general absorption scheme by reinterpreting $\kappa_0$ as the Laplace variable conjugate to the occupation time $a$ of the propagator $\calP(x,a,t)$ with the initial condition $\calP(x,a,t)=\delta(x-x_0)\delta(a)$. In particular, from Eq. (\ref{px0}),
\begin{widetext}
\begin{eqnarray}
 \PP(0,a,s)&=&  \int_0^{\infty}\e^{-st}\calP(0,a,t)dt ={\mathcal L}^{-1}_a \left [\frac{\cosh(\sqrt{s/D}(L-x_0))}{\sqrt{sD}\sinh(\sqrt{s/D}L)+\nu(s+z)\cosh(\sqrt{s/D}L)}\right ]\nonumber \\
 &=&{\mathcal L}^{-1}_a \left [\frac{\cosh(\sqrt{s/D}(L-x_0))}{\nu \cosh(\sqrt{s/D}L)}\frac{1}{z+s+\nu^{-1}\sqrt{sD}\tanh(\sqrt{s/D}L)}\right ]\nonumber \\
 &=&\frac{\cosh(\sqrt{s/D}(L-x_0))}{\nu \cosh(\sqrt{s/D}L)}\exp\left (-sa-\nu^{-1}\sqrt{sD}\tanh(\sqrt{s/D}L)a\right ).
\label{PPS}
\end{eqnarray}

The survival probability at time $t$ is now
\begin{equation}
\label{SPpsi}
S^{\Psi}(x_0,t)=\int_0^{\infty}\Psi(a)\left (\int_0^{L}P(x,a,t)dx+Q(a,t)\right )da,
\end{equation}
and
\begin{eqnarray}
\frac{\partial S^{\Psi}(x_0,t)}{\partial t}&=&\int_0^{\infty}\Psi(a)\left (D\int_0^L\frac{\partial^2 P(x,a,t}{\partial x^2}dx+\frac{\partial Q(a,t)}{dt}\right )da 
 =\int_0^{\infty}\Psi(a)\left (-D\frac{\partial P(0,a,t)}{\partial x} +\frac{\partial Q(a,t)}{dt}\right )da\nonumber \\
 &=& \int_0^{\infty}\Psi(a)\left (\frac{\partial Q(a,t))}{\partial a}+\delta(a)Q(0,t)\right )da= \int_0^{\infty}\psi(a)  Q(a,t))da,
\end{eqnarray}
\end{widetext}
where $\psi(a)=-\Psi'(a)$, and we have used Eqs. (\ref{BVPa})--(\ref{BVPc}). It follows that
the generalized MFPT is 
\begin{eqnarray}
\label{MFPTPsi}
\tau^{\Psi}(x_0)&=-\lim_{s\rightarrow 0} \frac{\partial}{\partial s}\J^{\Psi}_0(x_0,s),
 \end{eqnarray}
 where $\J^{\Psi}_0(x_0,s)$ is the Laplace transform of the flux into the origin:
\begin{equation}
   \J_0^{\Psi}(x_0,s) : = \nu\int_0^{\infty}\psi(a) \widetilde{P}(0,a,s)da.
\end{equation}
Substituting for $\PP(0,a,s)$ using Eq. (\ref{PPS}) gives
\begin{eqnarray}
\label{JPsi}
  \J^{\Psi}_0(x_0,s)= \frac{\cosh(\sqrt{s/D}(L-x_0))}{  \cosh(\sqrt{s/D}L)}\widetilde{\psi}  (\Gamma(s)),
 \end{eqnarray}
 where $\widetilde{\psi}(z)$ is the Laplace transform of $\psi(a)$, and
 \begin{equation}
 \Gamma(s)=s+\nu^{-1}\sqrt{sD}\tanh(\sqrt{s/D}L).
 \end{equation}
 Taylor expanding $ \J^{\Psi}_0(x_0,s)$ with respect to $s$ for small $s$ gives
\begin{eqnarray}
  \J^{\Psi}_0(x_0,s)&=& \left (1-s\frac{L^2-(L-x_0)^2}{2D}\right )\\
  &&\quad \times
\left (\widetilde{\psi}(0)+s\left (1+\frac{L}{\nu} \right )\widetilde{\psi}'(0)\right )+h.o.t.\nonumber 
 \end{eqnarray}
 Hence,
\begin{eqnarray}
\label{MFPT2}
\tau^{\Psi}(x_0)=\frac{L^2-(L-x_0)^2}{2D}+\E[a]\left (1+\frac{L}{\nu}\right ),
 \end{eqnarray}
 where $\widetilde{\psi}(0)=1$ and
 \begin{equation}
 \E[a]:=\int_0^{\infty}a\psi(a)da=-\widetilde{\psi}'(0).
 \end{equation}
 A number of results follow from this.
 \medskip
 
 \noindent (i) In the limit $\nu\rightarrow \infty$ with $\E[a]<\infty$, we recover the classical result for the MFPT in an interval with a totally absorbing boundary at $x=0$.
 \medskip
 
 \noindent (ii) The MFPT is infinite for densities $\psi(a)$ that do not have finite first moments. An example of a non-exponential density with finite moments at all orders is the gamma distribution
\begin{equation}
\label{psigam}
\psi(a)=\frac{\gamma(\kappa_0 a)^{\mu-1}\e^{-\kappa_0 a}}{\Gamma(\mu)},\quad \widetilde{\psi}(z)=\left (\frac{\kappa_0}{\kappa_0+z}\right )^{\mu} 
\end{equation}
for $\mu >0$ and $\mu \neq 1$, where $\Gamma(\mu)$ is the gamma function 
\begin{equation}
\Gamma(\mu)=\int_0^{\infty}\e^{-t}t^{\mu-1}dt,\ \mu >0.
\end{equation}
(If $\mu=1$ then we recover the exponential distribution with constant reactivity $\kappa_0$.) For the gamma distribution the mean occupation time is $\E[a]=\mu/\kappa_0$. Another well-known density is the Pareto-II (Lomax) distribution
\begin{equation}
 \psi(a)=\frac{\kappa_0 \mu}{(1+\kappa_0 a)^{1+\mu}},\, \widetilde{\psi}(z)=\mu\left (\frac{z}{\kappa_0}\right )^{\mu}\e^{z/\kappa_0}\Gamma(-\mu,z/\kappa_0) 
 \end{equation}
for $\mu>0$, where $\Gamma(\mu,z)$ is the upper incomplete gamma function:
\begin{equation}
 \Gamma(\mu,z)=\int_z^{\infty}\e^{-t}t^{\mu-1}dt,\ \mu >0.
\end{equation}
Using the identity
$\Gamma(1-\mu,z)=-\mu \Gamma(-\mu,z) +z^{-\mu}\e^{-z}$,
it can be checked that $\widetilde{\psi}(0)=1$, whereas $\E[a]$ is only finite if $\mu>1$. In the latter case
\begin{equation}
\E[a]= \frac{\Gamma(\mu-1)\Gamma(2)}{\kappa_0 \Gamma(\mu)}=\frac{1}{\kappa_0 (\mu-1)}.
\end{equation}
The blow up of the moments when $\mu<1$ reflects the fact that the Paretto-II distribution has a long tail.
\medskip

\noindent (iii) The MFPT blows up in the non-sticky limit $\nu \rightarrow 0$ for finite $\E[a]$, since the boundary at $x=0$ becomes totally reflecting. In order to recover the Robin boundary condition, it is necessary to scale the density $\psi(a)$ so that $\E[a]\rightarrow 0$ with $\overline{\kappa}_0=\nu/\E[a]$ fixed. We then recover the MFPT for a non-sticky boundary \cite{Bressloff22a}, which is given by 
\begin{eqnarray}
\label{MFPT0}
\overline{\tau}^{\Psi}(x_0)=\frac{L^2-(L-x_0)^2}{2D}+\frac{L}{\overline{\kappa}_0}.
 \end{eqnarray}

\setcounter{equation}{0}
\section{Steady state concentration for multiple sticky particles}

So far we have focused on an encounter-based model for single-particle diffusion. However, one can also apply the theory to multiple diffusing particles provided that they are treated as independent \cite{Bressloff22b}. Again consider diffusion in the interval $[0,L]$ with a partially absorbing sticky boundary at $x=0$. The propagator $P(x,a,t)$ is reinterpreted as the concentration $U(x,a,t)$ of particles having an occupation time $A(t)=a$ and, for a given distribution $\Psi(a)$, the concentration $u(x,t)$ is
\begin{equation}
\label{uU}
u(x,t)=\int_0^{\infty} \Psi(a)U(x,a,t)da.
\end{equation}
Similarly, the number of particles at $x=0$ is
\begin{equation}
v(t)=\int_0^{\infty} \Psi(a)V(a,t)da,\quad V(a,t)=\nu P(0,a,t).
\end{equation}
One difference from the single particle case is that we can now include a flux source term at the end $x=L$. There then exists a non-trivial steady-state solution 
  \begin{equation}
  U^*(x,a)=\lim_{t\rightarrow \infty} U(x,a,t)=\lim_{s\rightarrow 0}s \widetilde{U}(x,a,s),
  \end{equation}
and $V^*(a)=\nu U^*(0,a)$.  We wish to determine whether or not there is a corresponding steady-state solution $(u^*(x),v^*)$ and if so, investigate the relaxation to the steady state.
  
 The multi-particle version of the propagator evolution Eqs. (\ref{BVPa})--(\ref{BVPc}) for $x\in [0,L]$ is 
 \begin{subequations}
\begin{eqnarray}
\label{uBVPa}
\frac{\partial  U(x,a,t)}{\partial t}&=&D\frac{\partial^2 U(x,a,t)}{\partial x^2},\quad x\in (0,L),\\
\label{uBVPb} D\partial_xU(L,a,t)&=&J_L\delta(a),\quad V(a,t)=\nu U(0,a,t),\\
D\partial_xU(0,a,t)&=& \left [\frac{\partial V(a,t))}{\partial a}+\frac{\partial V(a,t)}{\partial t}+\delta(a)V(0,t)\right ].\nonumber \\
\label{uBVPc}
 \end{eqnarray}
\end{subequations}
 The boundary condition at $x=L$ includes the Dirac delta function $\delta(a)$, since newly injected particles have never interacted with the sticky boundary. In contrast to the single-particle case, we assume that at time $t=0$ there are no particles in $[0,L]$. Laplace transforming with respect to $a$ and $t$ gives
\begin{subequations}
\label{m1D}
\begin{eqnarray}
 && D\frac{\partial^2 \calU(x,z,s)}{\partial x^2}-s\calU(x,z,s) =0,\ x\in (0,L),  \\
 && D\partial_x\calU(0,z,s) =\nu(s+z) \calU(0,z,s),\\ 
 &&D \partial_x\calU(L,z,s)=\frac{J_L}{s}. 
 \end{eqnarray}
\end{subequations}
The general solution is
\begin{eqnarray}
\label{mQ1D}
 \calU(x,z,s)&=& A(z,s) \cosh(\sqrt{s/D} [L-x]) \\
 && -\frac{1}{\sqrt{sD}}\frac{J_L}{s}\sinh(\sqrt{s/D} [L-x]) .
\end{eqnarray}
The unknown coefficient $A(z,s)$ is determined by the sticky boundary condition at $x=0$:
\begin{eqnarray}
\label{mAA}
&&A(z,s)\\
&&=  \frac{\displaystyle J_L}{\displaystyle s}\frac{  \nu(s+z)\sinh(\sqrt{s/D}L) /\sqrt{sD}+\cosh(\sqrt{s/D}L) }{\nu(s+z)\cosh(\sqrt{s/D}L)+\sqrt{sD}\sinh(\sqrt{s/D}L)}.\nonumber
\end{eqnarray}
Multiplying the solution (\ref{mQ1D}) by $s$ and taking the limit $s\rightarrow 0$ yields the non-trivial steady-state solution
\begin{equation}
\calU^*(x,z)=J_L\left [\frac{x}{D}+\frac{1}{\nu z }\right ],\quad {\mathcal V}^*(z):=\nu \calU^*(0,z)=\frac{J_L}{z}.
\end{equation}
Inverting with respect to $z$ then implies
\begin{equation}
U^*(x,a)=J_L\left [\frac{x}{D}\delta(a)+\frac{1}{\nu }\right ],\quad V^*(a)=J_L.
\end{equation}

It follows that the steady-state concentration $u^*(x)$ for a given threshold distribution $\Psi(a)$ is
\begin{equation}
u^*(x)=\int_0^{\infty}\Psi(a)U^*(x,a)da= J_L\left [\frac{x}{D}+\frac{\E[a]}{\nu}\right ],
\label{uss}
\end{equation}
since $\widetilde{\Psi}(0)=\E[a]$. Similarly, the steady-state number of particles at $x=0$ is $v^*= J_L\E[a]$. Hence, a steady-state solution will only exist if $\E[a]<\infty$. 
 
When $u^*(x)$ exists we can quantify the rate of relaxation in terms of the so-called accumulation time, following along analogous lines to our previous analysis of a non-sticky boundary \cite{Bressloff22b}. 
The accumulation time was originally introduced within the context of diffusion-based protein concentration gradient formation \cite{Berez10,Berez11,Bressloff19}. We begin by briefly recalling the definition of the accumulation time. (For further details and comparisons with standard spectral analysis see Ref. \cite{Bressloff22b}.) Introduce the function
\begin{equation}
\label{accu}
Z(x,t)=1-\frac{u(x,t)}{u^*(x)},
\end{equation}
which is the fractional deviation of the concentration from steady state. Assuming that there is no overshooting, $1-Z(x,t)$ can be interpreted as the fraction of the steady-state concentration that has accumulated at $x$ by time $t$. It follows that $-\partial_t Z(x,t)dt$ is the fraction accumulated in the interval $[t,t+dt]$. The accumulation time $T_{\rm acc}(x)$ at position $x$ is then defined as 
\begin{equation}
\label{accu2}
T_{\rm acc}(x)=\int_0^{\infty} t\left (-\frac{\partial Z(x,t)}{\partial t}\right )dt=\int_0^{\infty} Z(x,t)dt.
\end{equation}
The Laplace transform of Eq. (\ref{accu}) gives
\[s\widetilde{Z}(x,s)=1-\frac{s\widetilde{u}(x,s)}{u^*(x)}\]
and, hence
\begin{eqnarray}
  T_{\rm acc}(x)&=&\lim_{s\rightarrow 0} \widetilde{Z}(x,s) = \lim_{s\rightarrow 0}\frac{1}{s}\left [1-\frac{s\widetilde{u}(x,s)}{u^*(x)}\right ] \nonumber\\
  &=&-\frac{1}{u^*(x)}
\left .\frac{d}{ds}[s\widetilde{u}(x,s)]\right |_{s=0}.
\label{acc}
\end{eqnarray}
We have used the result
\[u^*(x)=\lim_{t\rightarrow \infty} u(x,t)=\lim_{s\rightarrow 0}s\widetilde{u}(x,s).\]

\begin{figure}[t!]
\centering
\includegraphics[width=8cm]{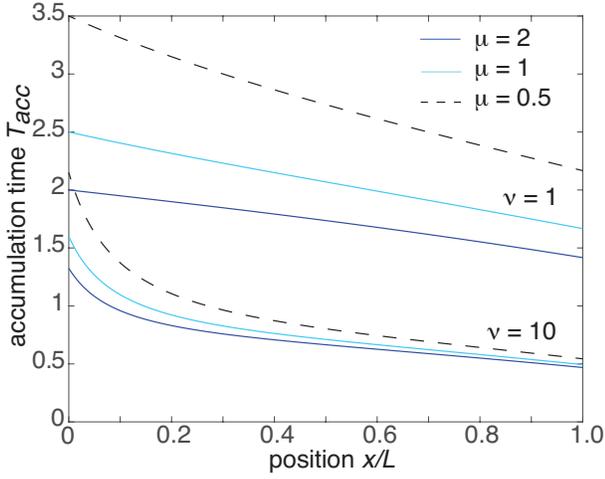}
\caption{Accumulation time $T_{\rm acc}(x)$ for diffusion in the interval $[0,L]$ with a partially absorbing sticky boundary at $x=0$ and a constant influx at $x=L$. We plot $T_{\rm acc}(x)$, see Eq.   (\ref{tact2}), as a function of $x$ for $\nu=1$ (upper curves) and $\nu=10$ (lower curves) in the case of the gamma distribution (\ref{psigam}). We fixed $\E[a]=1$ so that $\E[a^2]=1+1/\mu$ and consider different choices for $\mu$. We also set $L=D=1$.}
\label{fig4}
\end{figure}

\begin{figure}[t!]
\centering
\includegraphics[width=8cm]{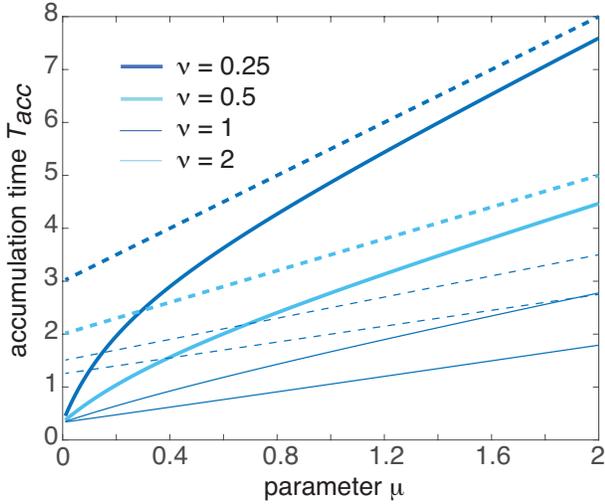}
\caption{Plot of $T_{\rm acc}(x)$ as a function of the gamma distribution parameter $\mu$ for $x=1$ (solid curves) and $x=0$ (dashed curves). We also set $\kappa_0=L=D=1$.}
\label{fig5}
\end{figure}

Eq. (\ref{uU}) implies that in order to calculate the accumulation time $T_{\rm acc}(x)$, we first have to determine the small-$s$ behavior of $s\widetilde{U}(x,z,s)$ and then invert with respect to $z$.
The simplest approach is to Taylor expand the solution (\ref{mQ1D}) with respect to $s$. After some algebra we find that
\begin{eqnarray}
 -\left .\frac{d}{ds}s\calU(x,z,s)\right |_{s=0}=J_L\left (c_0(x)+z^{-1}c_1(x)+z^{-2}c_2(x)\right ),\nonumber \\
\end{eqnarray}
with
\begin{eqnarray}
c_0(x)&=&\frac{L^3}{3D^2}-\frac{[L-x]^2L}{2D^2}+\frac{[L-x]^3}{6D^2}, \\
 c_1(x)&=& \frac{2L^2-[L-x]^2}{2\nu D},\ c_2(x)=\frac{1}{\nu}\left (1+\frac{L}{\nu}\right ).\nonumber
  \label{cx}
\end{eqnarray}
Inverting with respect to $z$ implies that
\begin{eqnarray}
  -\left .\frac{d}{ds}s\U(x,a,s)\right |_{s=0}=J_L\left (c_0(x)\delta(a)+c_1(x)+c_2(x)a\right ).\nonumber\\
  \label{noo}
\end{eqnarray}
Finally, multiplying both sides of the previous equation by $\Psi(a)/u^*(x)$ and integrating with respect to $a$ leads to the following explicit expression for the accumulation time:
\begin{eqnarray}
 T_{\rm acc}(x)&=&\frac{J_L}{u^*(x)} \bigg (c_0(x)+c_1(x)\int_0^{\infty}\Psi(a)da\nonumber \\
 &&\quad +c_2(x)\int_0^{\infty}a \Psi(a)da\bigg).
 \label{tact}
\end{eqnarray}
Moreover, $\int_0^{\infty}\Psi(a)da=\widetilde{\Psi}(0)=\E[a]$ and
\begin{eqnarray}
 \int_0^{\infty} a\Psi(a)da&=&\frac{1}{2}[a^2 \Psi(a)]_0^{\infty}-\frac{1}{2}\int_0^{\infty}a^2 \Psi'(a)da \nonumber \\
 &=& \frac{1}{2}\int_0^{\infty}a^2\psi(a)da=\frac{1}{2}\E[a^2],
\end{eqnarray}
so that
\begin{eqnarray}
\label{tact2}
 T_{\rm acc}(x)=\frac{J_L}{u^*(x)} \left (c_0(x)+c_1(x)\E[a]+\frac{c_2(x)}{2}\E[a^2]\right).\nonumber \\
\end{eqnarray}
Note that $T_{\rm acc}(x)$ is independent of the flux $J_L$.
Eq. (\ref{tact2}) also implies that the accumulation time is only well-defined if the first and second moments of $\psi(a)$ are finite. In the particular case of the gamma distribution (\ref{psigam}), we have
\begin{equation}
\E[a] =\frac{\mu}{\kappa_0},\quad \E[a^2]=\frac{\mu(\mu+1)}{\kappa_0^2}.
\end{equation}

\begin{figure}[t!]
\centering
\includegraphics[width=8cm]{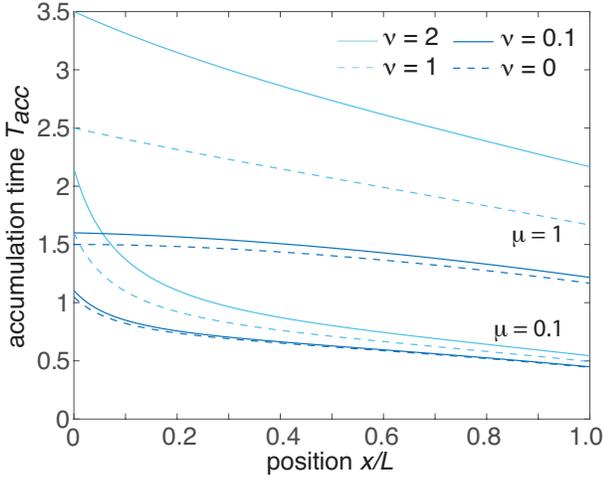}
\caption{Plot of $T_{\rm acc}(x)$ as a function of $x$ for $\mu=1$ (upper curves) and $\mu=0.1$ (lower curves) with $\overline{\kappa_0}:=\kappa_0\nu=1$. In the limit $\nu\rightarrow 0$ and $\kappa_0\rightarrow \infty$, the accumulation curves converge to the case of a partially absorbing non-sticky boundary.We also set $L=D=1$.}
\label{fig6}
\end{figure}

\begin{figure}[t!]
\centering
\includegraphics[width=8cm]{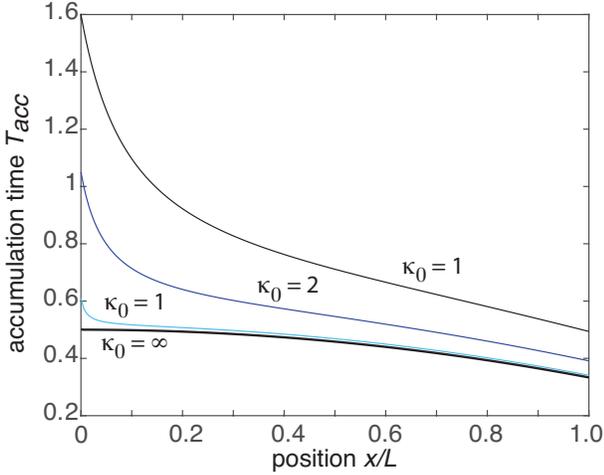}
\caption{Plot of $T_{\rm acc}(x)$ as a function of $x$ for $\mu=0.1$, $\nu=1$ and various values of $\kappa_0$. In the limit $\kappa_0\rightarrow \infty$, the accumulation curves converge to the case of a totally absorbing boundary, which is independent of $\nu$ and $\mu$. We also set $L=D=1$.}
\label{fig7}
\end{figure}

Given the more complicated dependence of the accumulation time on model parameters compared to the MFPT, we illustrate its behavior in Figs. \ref{fig4}--\ref{fig7} for the gamma distribution. First, Fig. \ref{fig4} shows plots of $T_{\rm acc}(x)$ against $x$ for different values of $\nu$ and $\E[a]=1$. We see that increasing $\nu$ reduces the accumulation time for all $x$ but leads to a sharper increase in $T_{\rm acc}(x)$ as $x\rightarrow 0$. The latter reflects the greater stickiness of the boundary. The accumulation time is also an increasing function of the $\mu$ for fixed $\kappa_0$, which is further illustrated in Fig. \ref{fig5}.
It can also be checked that our previous results for a non-sticky boundary are recovered in the limit $\nu\rightarrow 0$ and $\kappa_0\rightarrow \infty$ with $\overline{\kappa}_0=\nu \kappa_0$ fixed \cite{Bressloff22b}. This is shown in Fig. \ref{fig6} where we plot $T_{\rm acc}(x)$ as a function of $x$ for fixed $\overline{\kappa}_0$ and different stickiness parameter values. Finally, in the limit $\kappa_0\rightarrow \infty$, the boundary become totally absorbing irrespective of the level of stickiness.  This is illustrated in Fig. \ref{fig7}.

We end by noting that we can also define an accumulation time for the number of particles at $x=0$, namely,
\begin{equation}
\tau_{\rm acc}=\int_0^{\infty} \left [1-\frac{v(t)}{v^*}\right ]dt.
\end{equation}
In terms of Laplace transforms, this becomes
\begin{eqnarray}
   \tau_{\rm acc} =-\frac{1}{v^*}
\left .\frac{d}{ds}[s\widetilde{v}(s)]\right |_{s=0},
\end{eqnarray}
with
\begin{eqnarray}  
\widetilde{v}(s)= \nu\int_0^{\infty} \Psi(a)\widetilde{U}(0,a,s)da. 
\end{eqnarray}
Eqs. (\ref{cx}) and (\ref{noo}) yields the result
\begin{eqnarray}
 \tau_{\rm acc} &=&T_{\rm acc}(0)=\frac{\nu}{ \E[a]} \left ( \frac{L^2}{2\nu D}\E[a]+\frac{1}{2\nu}\left (1+\frac{L}{\nu}\right )\E[a^2]\right).\nonumber
\\
\label{tactt}
\end{eqnarray}

\setcounter{equation}{0}
\section{Sticky Brownian motion as the diffusion limit of a velocity jump process}

Another class of model where sticky boundary conditions are used is an unbiased two-state velocity jump process describing the motion of a run-and-tumble particle (RTP) \cite{Angelani17,Bressloff23}. Although the RTP model in an unbounded domain has a well-defined diffusion limit, it is not immediately clear how to perform the diffusion limit in the presence of a sticky boundary. In this section we show how this can be achieved using an analog of the localized potential energy function of Sect. IIB. We begin by considering the simpler case of an RTP in $\R$, and then turn to the case of a sticky boundary.

\subsection{Diffusion limit of run-and-tumble dynamics in $\R$} 
Let $X(t)\in \R$ denote the position of a particle at time $t$. Suppose that the particle switches between  two constant velocity states labeled by $\sigma=\pm $ with $v_{\sigma}=v_0\delta_{\sigma,+}-v_0\delta_{\sigma,-}$ and $v_0>0$.  Assuming that the particle reverses direction according to a Poisson process with rate $\alpha$, the dynamics of $X(t)$ for $X(t)>0$ is given by the piecewise deterministic equation
\begin{equation}
\label{PDMP}
\frac{dX}{dt}=v_{\sigma(t)},
 \end{equation}
where $\sigma(t)$ is a dichotomous noise process. Let $p_{\sigma}(x,t)$ be the probability density that at time $t$ the particle is at $X(t)=x$ and in the discrete state $\sigma(t)=\sigma=\pm$. The pair $p_{\pm}(x,t)$ evolve according to the differential Chapman-Kolmogorov (CK) equation
\begin{subequations}
\begin{eqnarray}
\label{DLa}
\frac{\partial p_{+}}{\partial t}&=&-v_0 \frac{\partial p_{+}}{\partial x}-\alpha p_{+}+\alpha p_{-},\\
\frac{\partial p_{-}}{\partial t}&=&v_0 \frac{\partial p_{-}}{\partial x}-\alpha p_{-}+\alpha p_{+}.
\label{DLb}
\end{eqnarray}
\end{subequations}

Equations of the form (\ref{DLa}) and (\ref{DLb}) were originally introduced by Goldstein \cite{Goldstein51} and Kac \cite{Kac74} to describe a correlated random walk. More recently, two-state velocity jump processes have arisen in a wide range of applications to cell biology. For example, $X(t)$ could represent the position of a bacterium confined to a semi-infinite channel with the end $x=0$ corresponding to the channel wall \cite{Angelani15,Angelani17}. Indeed the term ``run-and-tumble'' is based on the observed motion of  bacteria such as {\em E. coli} \cite{Berg04}. Alternatively, $X(t)$ could represent the tip of a growing and shrinking polymer filament that nucleates at the end $x=0$ \cite{Dogterom93,Gopa11,Mulder12,Bressloff19}. In order to establish that the RTP model has a well defined diffusion limit \cite{Kac74,Schnitzer93,Hillen00}, consider the change of variables
\begin{equation}
\label{cov}
 p =p_+ +p_- ,\quad J =v_0 [p_+ -p_- ],
\end{equation}
with $p$ the marginal probability for particle position and $J$ the probability flux. Eqs. (\ref{DLa}) and (\ref{DLb}) can then be rewritten in the form
\begin{eqnarray}
\label{JDL}
\frac{\partial p}{\partial t}=-\frac{\partial J}{\partial x},\quad
\frac{\partial p}{\partial t}= -v_0^2\frac{\partial p}{\partial x}-2\alpha J.
\end{eqnarray}
Suppose that we fix the length and time units so that $v_0$ and $\alpha$ are $O(1)$. Under the change of variables $T=\delta^2 t$ and $X= \delta x$, with $\delta $ a small dimensionless constant, Eqs. (\ref{JDL}) become
\begin{equation}
\label{scale}
\delta^2\frac{\partial p}{\partial T}=-\delta \frac{\partial J}{\partial X},\quad \delta^2\frac{\partial J}{\partial T}= - v_0^2\delta\frac{\partial p}{\partial X}-2\alpha J.
\end{equation}
Taking the diffusion limit $\delta \rightarrow 0$ is essentially equivalent to ignoring short-time transients. These equations can be balanced by taking $J=O(\delta)$ and dropping the $\partial J\partial t$ term. Hence, setting $p(X/\delta,T/\delta^2)=\rho(X,T)$ and $J(X/\delta,T/\delta^2)= j(X,T)\delta$ we obtain the leading order pair of equations \cite{Kac74}
\begin{equation}
\frac{\partial \rho}{\partial T}=-\frac{\partial j}{\partial X}, \quad v_0^2 \frac{\partial \rho}{\partial X}=2\alpha j.
\end{equation} 
Hence, $\rho(X,T)$ satisfies the diffusion equation
\begin{equation}
\frac{\partial \rho}{\partial T}=D\frac{\partial^2\rho }{\partial X^2} \mbox{ with } D=\frac{v_0^2}{2\alpha}.
\end{equation}
(Note that if we were interested in both the short-time and long-time dynamics then we would need to use the method of multiple scales.)

Another way to understand the diffusion limit is to observe that by differentiating Eqs. (\ref{DLa}) and  (\ref{DLb}) one finds that the marginal probability density $p(x,t)$ satisfies the telegrapher's equation   \cite{Goldstein51}
\begin{equation}
\label{tel}
\left [\frac{\partial^2}{\partial t^2}+2\alpha \frac{\partial}{\partial t}-v_0^2\frac{\partial^2}{\partial x^2}\right ]p(x,t)=0.
\end{equation}
Solutions of the telegrapher's equation generally exhibit distinct short-time and long-time behaviors. At short times (for $t\ll \tau_c=1/2\alpha$) the evolution is characterized by wave-like propagation with $\langle x(t)\rangle^2\sim (v_0t)^2$, whereas at long times ($t\gg 
\tau_c$) the behavior diffusive with $\langle x^2(t)\rangle \sim 2Dt,\, D=v_0^2/2\alpha$. As an explicit example, the solution on $\R$ for the initial conditions 
$p(x,0)=\delta(x)$ and $\partial_tp(x,0)=0$ is 
\begin{widetext}
\begin{eqnarray}
 p(x,t)&=&\frac{e^{-\alpha t}}{2}[\delta(x-v_0t)+\delta(x+v_0t)]+\frac{\alpha\e^{-\alpha t}}{2v} \bigg [I_0(\alpha\sqrt{t^2-x^2/v_0^2})+\frac{t}{\sqrt{t^2-x^2/v_0^2}}I_0(\alpha\sqrt{t^2-x^2/v_0^2})\bigg ]\nonumber\\
 &&\times[\Theta(x+v_0t)-\Theta(x-v_0t)],
\end{eqnarray}
\end{widetext}
where $I_n$ is the modified Bessel function of $n$-th order and $\Theta$ is the Heaviside function. The first two terms clearly represent the ballistic propagation of the 
initial data along characteristics $x=\pm v_0t$, whereas the Bessel function terms asymptotically approach Gaussians in the large time limit.

\subsection{Diffusion limit of a sticky boundary at $x=0$} 

\begin{figure}[b!]
\centering
\includegraphics[width=8cm]{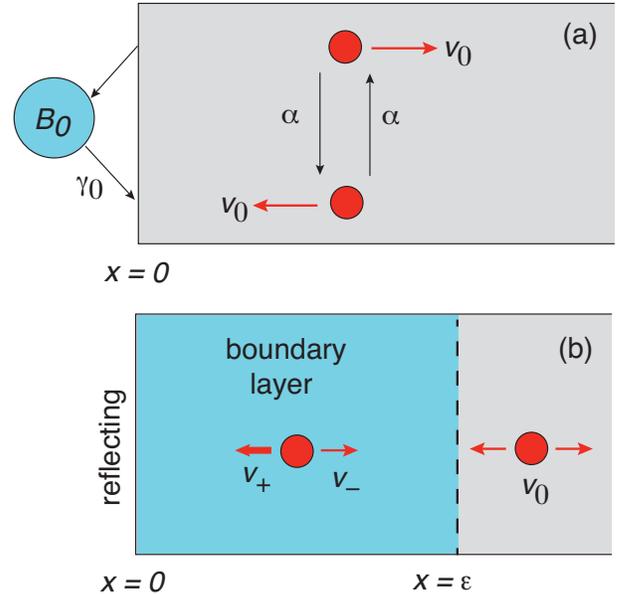}
\caption{Two implementations of a sticky boundary for run-and-tumble dynamics. (a) The RTP switches between the velocity states $\pm v_0$ at a Poisson rate $\alpha$. When the particle hits the end $x=0$ it enters the bound state $B_0$ until reentering the bulk domain at a constant rate $\gamma_0$. (b) Replacement of a bound state $B_0$ by a boundary layer of width $\epsilon$ at $x=0$ within which the velocity is spatially varying and biased.}
\label{fig8}
\end{figure}

Now suppose that there is a sticky boundary at $x=0$ \cite{Angelani17,Bressloff23}, see Fig. \ref{fig8}(a). Whenever the RTP hits the boundary, its velocity immediately drops to zero and it sticks to the boundary until it reverses its velocity state at a Poisson rate $\gamma_0$. (Note that $\gamma_0$ can differ from the switching rate $\alpha$ in the bulk.) 
Let $Q_0(t)$ be the probability that at time $t$ the particle is in the bound state $B_0$. The sticky boundary conditions at $x=0$ take the form
\begin{subequations}
\begin{equation}
\label{mstick0}
\gamma_0 Q_0(t)=vp_+(0,t)=\frac{vp(0,t)+J(0,t)}{2},
\end{equation}
with $Q_0(t)$ evolving according to the equation
\begin{equation}
\label{stick0}
\frac{dQ_0}{dt}=vp_-(0,t)-\gamma_0  Q_0(t)=-J(0,t).
\end{equation}
\end{subequations}
Conservation of probability implies that
\begin{equation}
\label{norm}
\int_0^{\infty}p(x,t)dx+Q_0(t)=1,\quad t>0.
\end{equation}
If $\gamma_0=0$, then the boundary $x=0$ is totally absorbing in the sense that, when the RTP hits the boundary, it irreversibly enters the bound state $B_0$ and $vp_+(0,t)=0$. On the other hand, if $v$ is finite then the boundary becomes totally reflecting in the limit $\gamma_0\rightarrow \infty$, since $Q_0(t)=0$ and hence $J(0,t)=0$. If we apply the change of variables $T= \delta^2 t$ and $X=\delta x$ with $\gamma_0=O(1)$, then we obtain Eqs. (\ref{scale}) together with the sticky boundary conditions 
\begin{equation}
\label{Q0}
Q_0(t)=\frac{ v_0p(0,t)+J(0,t)}{2\gamma_0},\quad \delta^2 \frac{dQ_0}{dt}=-J(0,t).
\end{equation}
Unfortunately, it is not possible to balance these equations for $\gamma_0 < \infty$. That is, the first equation   implies that $Q_0(t)=O(1)$ so that the second equation requires $J(0,t)=O(\delta^2)$, which differs from the bulk condition $J(x,t)=O(\delta)$. The latter could be resolved by taking $\gamma_0=O(\delta)$ but then $Q_0(t)=O(1/\delta)$, which blows up in the diffusion limit. The only consistent case is $\gamma_0\rightarrow \infty$, which corresponds to a totally reflecting boundary.

Therefore, motivated by the construction in Sect. IIB, we look for an alternative implementation of a sticky RTP based on the introduction of a boundary layer, see Fig. \ref{fig8}(b). A natural analog of a localized attractive potential energy function is a localized spatial variation in the velocity states that biases the motion so the boundary is attractive. That is, we take $v_{\pm}=v_0\mp \phi^{\epsilon}(x)$ with $\phi^{\epsilon}(x)\approx 0$ for $x\geq \epsilon$. Eqs. (\ref{DLa}) and (\ref{DLb}) become
\begin{subequations}
\begin{eqnarray}
\label{bDLa}
\frac{\partial p_{+}}{\partial t}&=&-\frac{\partial [v_0-\phi^{\epsilon}(x)] p_{+}}{\partial x}-\alpha p_{+}+\alpha p_{-},\\
\frac{\partial p_{-}}{\partial t}&=&\frac{\partial[v_0+\phi^{\epsilon}(x)] p_{-}}{\partial x}-\alpha p_{-}+\alpha p_{+},
\label{bDLb}
\end{eqnarray}
\end{subequations}
together with the reflecting boundary condition
\begin{equation}
v_+(0)p_+(0,t)-v_-(0)p_-(0,t)\equiv J(0,t)-\phi^{\epsilon}(0)p(0,t)=0.
\end{equation}
Adding and subtracting these equations yields
\begin{subequations}
\label{bJDL}
\begin{eqnarray}
\frac{\partial p}{\partial t}&=&-\frac{\partial J}{\partial x}+\frac{\partial \phi^{\epsilon}(x)p}{\partial x},\\
\frac{\partial J}{\partial t}&=& -v_0^2\frac{\partial p}{\partial x}+\frac{\partial \phi^{\epsilon}(x)J}{\partial x}-2\alpha J.
\end{eqnarray}
\end{subequations}

The main issue is whether or not it is possible to find a velocity function $\phi^{\epsilon}$ that leads to sticky BM in the diffusion limit. It turns out that the answer depends on which order we take the dual limits $\epsilon \rightarrow 0$ and $\delta \rightarrow 0$. Suppose that we take the limit $\epsilon \rightarrow 0$ first and perform matched asymptotics along analogous lines to Sect. IIB.
Let $(p,J)$ denote the leading order components of the asymptotic expansion of the outer solution in $x\geq \epsilon$ with
\begin{eqnarray}
\label{outerRTP}
\frac{\partial p}{\partial t}=-\frac{\partial J}{\partial x},\quad
\frac{\partial J}{\partial t}= -v_0^2\frac{\partial p}{\partial x}-2\alpha J.
\end{eqnarray}
Similarly, let $(q,K)$ be the corresponding leading order components of the inner solution within the boundary layer. Introducing the stretched coordinate $X=x/\epsilon$ and keeping only the leading order terms gives
\begin{subequations}
\begin{eqnarray}
\label{ka}
0&=& -\frac{\partial K(\epsilon X,t)}{\partial X}+\frac{\partial}{\partial X}\left [ \phi^{\epsilon}(\epsilon X)q(\epsilon X,t)\right ]\\
0&=&-v_0^2\frac{\partial q(\epsilon X,t)}{\partial X}+\frac{\partial \phi^{\epsilon}(\epsilon X)K(\epsilon X,t)}{\partial X},
\label{kb}
\end{eqnarray}
\end{subequations}
for $X\in (0,1)$ and $K(0,t)-\phi^{\epsilon}(0)q(0,t)=0$.

Eq. (\ref{ka}) and the reflecting boundary condition imply that
\begin{equation}
K(\epsilon X,t)=\phi^{\epsilon}(\epsilon X)q(\epsilon X,t),\quad X\leq 1.
\end{equation}
On the other hand, Eq. (\ref{kb}) gives
\begin{equation}
v_0^2q(\epsilon X,t) -\phi^{\epsilon}(\epsilon X)K(\epsilon X,t)=v_0^2C(t).
\end{equation}
for some function $C(t)$.
Combining the last two equations, we have
\begin{equation}
q(\epsilon X,t)=\frac{v_0^2C(t)}{v_0^2-[\phi^{\epsilon}(\epsilon X)]^2} .
\end{equation}
In order to obtain to determine $C(t)$ we first impose continuity of the solution at $x=\epsilon$ and $X=1$, which yields $C(t)= p(0,t)$ to leading order. (Recall the assumption that $\phi^{\epsilon}(x)\approx 0$ for $x\geq \epsilon$.) Conservation of probability then implies that
\begin{eqnarray}
0&=&\frac{d}{\partial t}\left (\int_0^{\epsilon}q(x,t)dx+\int_{\epsilon}^{\infty} p(x,t)dx\right ) ,\nonumber \\
0&= &\frac{dC(t)}{dt}\int_0^{\epsilon}\frac{v_0^2dx}{v_0^2-[\phi^{\epsilon}(x)]^2} +J(\epsilon,t).
\label{rtpa3}
\end{eqnarray}
Analogous to Eq. (\ref{a3}) we impose the condition
\begin{equation}
\lim_{\epsilon \rightarrow 0} \int_0^{\epsilon}\frac{v_0^2dx}{v_0^2-[\phi^{\epsilon}(x)]^2}=\nu .
\end{equation}
 Taking the limit $\epsilon \rightarrow 0$ in Eq. (\ref{rtpa3}) and setting $Q_0(t)=\nu p(0,t)$ recovers the boundary condition (\ref{stick0}). Moreover, Eq. (\ref{mstick0}) allows us to relate the stickiness parameter $\nu$ to the rate $\gamma_0$:
 \begin{equation}
 \nu=\frac{v_0+\phi^{\epsilon}(0)}{2\gamma_0}.
 \end{equation}
Finally, the conservation condition follows from noting that in the limit $\epsilon \rightarrow 0$, the solution of Eqs. (\ref{bJDL}) is formally given by $p(x,t)(1+\nu \delta(x))$. As we have already pointed out, it is not possible to obtain sticky BM by taking the diffusion limit after taking $\epsilon \rightarrow 0$.

\begin{figure*}[t!]
\centering
\includegraphics[width=16cm]{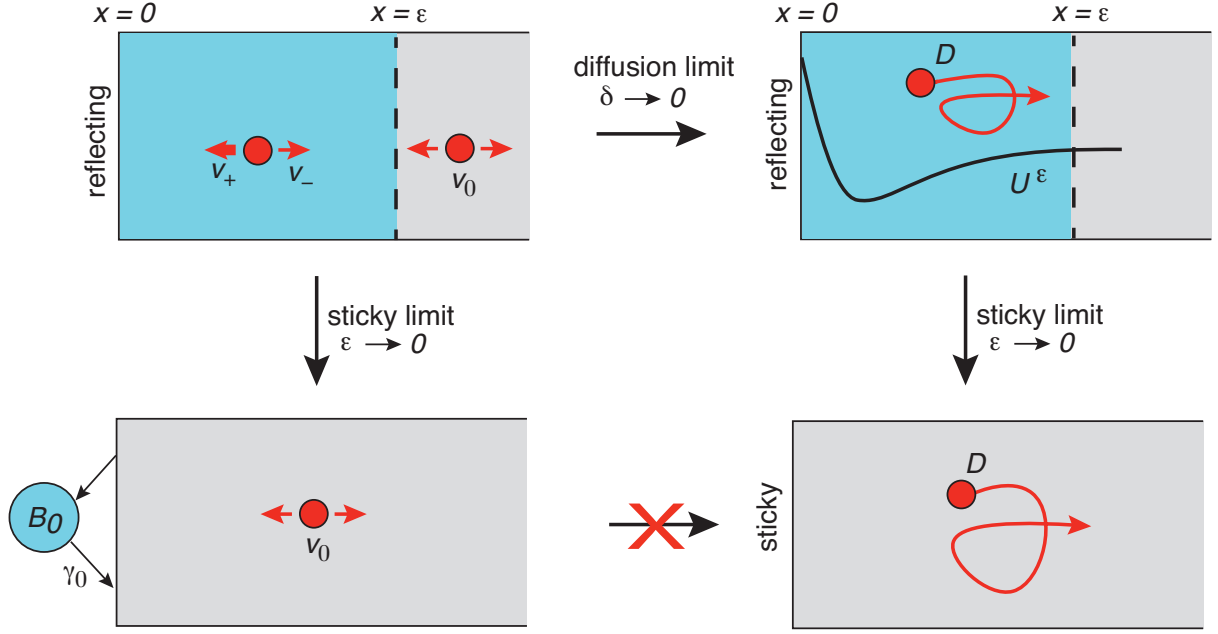}
\caption{Summary of the relationship between sticky BM and sticky run-and-tumble dynamics. Using two different choices of the velocity function $\phi^{\epsilon}(x)$ for reflecting RTP, we can either take the sticky limit $\epsilon \rightarrow 0$ for fixed $\delta$ or the diffusion limit $\delta \rightarrow$ for fixed $\epsilon$ to obtain, respectively, sticky RTP or reflecting BM with a local attractive potential. In the latter case, sticky BM is now obtained by taking the limit $\epsilon \rightarrow 0$. There is no direct path from sticky RTP to sticky BM.}
\label{fig9}
\end{figure*}

Therefore, suppose that we now take the diffusion limit for fixed $\epsilon$.
We also assume that under the change of variables $T= \delta^2 t$ and $X=\delta x$, the velocity function scales as
$\phi^{\epsilon}(X/\delta)= \Phi^{\epsilon}(X)\delta$. It follows that
\begin{subequations}
\label{bscale}
\begin{eqnarray}
 \delta^2\frac{\partial p}{\partial T}&=&-\delta \frac{\partial J}{\partial X}+\delta^2 \frac{\partial \Phi^{\epsilon}(X)p}{\partial X},\\ \delta^2\frac{\partial J}{\partial T}&=& - v_0^2\delta\frac{\partial p}{\partial X}+\delta^2 \frac{\partial \Phi^{\epsilon}(x)J}{\partial X}-2\alpha J.
\end{eqnarray}
\end{subequations}
These equations can be balanced for fixed $\epsilon >0$ by taking $J=O(\delta)$ and dropping the terms $\partial J/\partial T$ and ${\partial [\Phi^{\epsilon}(X)J]}/{\partial X}$. Hence, setting $p(X/\delta,T/\delta^2)=\rho(X,T)$, $J(X/\delta,t/\delta^2)= j(X,T)\delta$ and eliminating $j$ results in the FP equation
\begin{equation}
\frac{\partial \rho}{\partial T}=D\frac{\partial^2 \rho }{\partial X^2}+\frac{\partial \Phi^{\epsilon}(X)\rho}{\partial X},
\end{equation}
together with the reflecting boundary condition $D\partial_X\rho(0,T)+\Phi^{\epsilon}(X)\rho(0,T)=0$.
Comparison with Eqs. (\ref{qFP}) and (\ref{qBC}) shows that in the diffusion limit for fixed $\epsilon$, we recover the boundary-layer construction of sticky BM under the identification $\Phi^{\epsilon}(x)=\partial_xU^{\epsilon}(x)$. We conclude that the choice of velocity function plays a crucial role in determining the appropriate sticky limit. Our results regarding the relationship between sticky BM and sticky run-and-tumble dynamics are summarized in Fig. \ref{fig9}.

\subsection{Encounter-based model of sticky run-and-tumble dynamics}

For the sake of completeness, we briefly summarize the encounter-based model of sticky RTP. For further details, see Ref. \cite{Bressloff23}. First, we extend the discrete state space of the RTP by taking $\sigma(t)\in \{\pm,0\}$ with $\sigma(t)=0$ if the RTP is in the bound state $B_0$. Let $A_0(t)$ denote the amount of time that the RTP spends in state $B_0$ over the time interval $(0,t)$. 
Introduce the joint probability density or occupation time propagator
\begin{eqnarray}
 &&\calP_{\sigma}(x,a,t)dxda \\
 &&=\P[x<X(t)<x+dx,a<A_0(t)<a+da,\sigma(t)=\sigma],\nonumber
\end{eqnarray}
with $X(0)=x_0$, $\sigma(0)=\pm$ with probability 1/2, and $A(0)=0$.
Since the occupation time $A_0(t)$ only changes at the boundaries, the evolution equation within the bulk of the domain takes the same form as for $p_{\sigma}(x,t)$, see Eqs. (\ref{DLa}) and (\ref{DLb}). In terms of the transformed propagators
\begin{subequations}
\begin{eqnarray}
\calP(x,a,t)&=&\calP_+(x,a,t)+\calP_-(x,a,t),\\ \calJ(x,a,t)&=&v_0[\calP_+(x,a,t)-\calP_-(x,a,t)],
\end{eqnarray}
\end{subequations}
we have
 \begin{eqnarray} 
   \label{JPCK}
   \frac{\partial \calP}{\partial t}   =- \frac{\partial \calJ}{\partial x},\quad
 \frac{\partial \calJ}{\partial t}   =-v_0^2 \frac{\partial \calP}{\partial x}-2\alpha \calJ .
  \end{eqnarray}
Similarly, defining
   \begin{equation}
   \calQ_{0}(a,t)da=\P[a<A_0(t)<a+da , \sigma(t)=0],
   \label{Qdef}
   \end{equation}
the boundary condition for the propagator can be expressed as
\begin{equation}
\label{calstick0}
 \frac{\partial \calQ_0}{\partial a}+\frac{\partial \calQ_0}{\partial t}=-\delta(a)\calQ_0(0,t)-\calJ(0,a,t), 
\end{equation}
with
\begin{equation}
 \calQ_0(a,t)=\frac{ v_0\calP(0,a,t)+\calJ(0,a,t)}{2\gamma_0}.
\end{equation}
The form of Eq. (\ref{calstick0}) is similar to Eq. (\ref{BVPc}). Hence, the sum of time derivatives on the left-hand side of Eq. (\ref{calstick0}) reflects the fact that whenever the RTP is in a bound state, the occupation time $A_0(t)$ increases at the same rate as the absolute time $t$. Moreover, the term involving the Dirac delta function $\delta(a)$ on the right-hand of Eqs. (\ref{calstick0}) ensures that the probability of being stuck at a boundary is zero if the occupation time is zero.
Analogous to the encounter-based model of diffusion, absorption can now be included by introducing a stopping time condition of the form (\ref{TA}) and setting
\begin{subequations}
\begin{eqnarray}
   p^{\Psi}(x,t)&=&\int_0^{\infty}\Psi(a) \calP(x,a,t)da  ,\\
 Q^{\psi}_{0}(t)&=&\int_0^{\infty}\Psi(a) \calQ_{0}(a,t)da  .
\end{eqnarray}
\end{subequations}
\smallskip

\section{Discussion} In this paper we derived the evolution equation for the occupation time propagator of sticky BM on the half-line, see Eqs. (\ref{BVPa})--(\ref{BVPc}). We achieved this by taking the sticky limit $\epsilon \rightarrow 0$ of the propagator equation for partially reflecting BM in the presence of a narrow and deep potential well within a boundary layer $[0,\epsilon]$. We then used the propagator to construct and analyze an encounter-based model of BM with a partially absorbing sticky boundary at the origin. We calculated the MFPT for absorption in a finite interval $[0,L]$ with a totally reflecting boundary at $x=L$. We also obtained the steady-state concentration in a multi-particle version of the model with a constant influx at $x=L$, and derived an explicit expression for the local rate of relaxation to the steady-state. We determined how these various quantities depended on the stickiness parameter $\nu$ and moments of the occupation time threshold density $\psi(a)$. Finally, we explored the relationship between sticky BM and the dynamics of a sticky RTP in the diffusion limit.

In order to develop the theory presented in this paper, we restricted our analysis to stochastic processes on the half-line. Indeed, most mathematical formulations of sticky BM have focused on this configuration. An obvious extension of our work would be to consider sticky BM in a $d$-dimensional domain $\Omega \subset \R^d$ with a $(d-1)$-dimensional boundary $\partial \Omega$ and $d>1$. One major difference from the 1D case is that the boundary is no longer a single point. This means that there could be a distinct form of motion on $\partial \Omega$. For example, in the case of cell polarization, where $\Omega$ would represent the cell interior and $\partial \Omega$ the cell membrane, one finds that diffusion in the membrane is significantly slower than diffusion in the cytoplasm \cite{Drake10,Jilkine11,Martin14,Halatek18}. There have been a few recent probabilistic formulations of multi-dimensional sticky BM \cite{Grothaus17,Aurell20}, but these have focused on mathematical features of solutions to the corresponding multi-dimensional SDEs. The approach taken in the current paper suggests considering higher-dimensional version of BM with a localized potential that allows for separate motion within $\partial \Omega$.

From a more general perspective, this paper has demonstrated how encounter-based models provide a powerful framework for studying stochastic processes in domains with reactive surfaces. Hence, it would be interesting in future work to expand the class of models to include anomalous diffusion processes such as fractional Brownian motion \cite{Metzler04,Metzler14}, and other examples of active particles \cite{Roman12}. For example, it is known that an active Brownian particle (ABP) in a two-dimensional channel also accumulates at a wall by pushing on the boundary until a tumble event reverses its direction away from the wall. However, the analysis of such a process is non-trivial, since the corresponding two-dimensional FP equation involves a so-called two-way diffusion problem. That is, the boundary conditions at the walls are only defined on orientation half spaces, which means that standard Sturm-Liouville theory cannot be applied \cite{Lee13,Wagner17,Wagner19}. We have recently developed an encounter-based model of an ABP, which simplifies the problem by assuming that the statistics of particle arrivals at the wall is known \cite{Bressloff23a}. This again exploits the separation of surface reactions from bulk motion.
Finally, another twist on encounter-based models is to replace a partially absorbing boundary by partially absorbing traps within the interior of the domain \cite{Bressloff22,Bressloff22a}. In this case a particle can freely enter and exit a trap, but can only be absorbed within the trapping region. The probability of absorption is taken to depend on the particle-trap encounter time which, in the case of BM, is given by the Brownian occupation time. The inclusion of some form of stickiness is less clear here, but one could imagine some mechanism that biases the motion so that it enhances the probability of lingering within the trapping region.


\begin{thebibliography}{9}
\bibitem{note2} One technical point is that there are no strong solutions for sticky BM $\widehat{Y}(t)$, meaning that not every realization of the Wiener process is mapped to a path of $\widehat{Y}(t)$. However, there does exist a weak solution in the sense that the probability density of sample paths (solution of the corresponding FP equation) is unique \cite{Engelbert14}.
\end{thebibliography}
\end{document}